\documentclass[pra,aps,amsmath,amssymb,amsfonts,twocolumn,nofootinbib,longbibliography]{revtex4-1}
\usepackage{}
\usepackage{amssymb}
\usepackage{bm,mathrsfs}
\usepackage{graphicx}
\usepackage{epsfig}
\usepackage{amsmath,bbm}
\usepackage{amsfonts,amssymb}
\usepackage{times}
\usepackage{verbatim}
\usepackage[sort&compress]{natbib}
\usepackage{amsmath}
\usepackage{bm}
\usepackage{float}
\allowdisplaybreaks[4]
\usepackage[colorlinks,breaklinks,linkcolor=blue,anchorcolor=blue,citecolor=blue,urlcolor=magenta]{hyperref}
\usepackage{fontawesome}

\usepackage{color}
\definecolor{zzz}{rgb}{0.9,0.0,0.4}

\begin{document}
\title{Nonreciprocal Quantum Router with Non-Markovian Environments}

\author{T. Z. Luan$^{1,2}$}

\author{Cheng Shang$^{3,4}$}
\email{Contact author: cheng.shang@riken.jp}

\author{H. Yi$^{1}$}

\author{J. L. Li$^{1}$}

\author{Yan-Hui Zhou$^{5}$}

\author{Shuang Xu$^{6}$}

\author{H. Z. Shen$^{1}$}
\email{Contact author: shenhz458@nenu.edu.cn}

\makeatletter
\renewcommand\frontmatter@affiliationfont{\vspace{1mm} \small}
\makeatother

\affiliation{\textit{
$^{1}$Center for Quantum Sciences and School of Physics, Northeast Normal University, Changchun 130024, China\\
$^{2}$School of Physics Science and Technology, Shenyang Normal University, Shenyang 110034, China\\
$^{3}$Analytical Quantum Complexity RIKEN Hakubi Research Team, RIKEN Center for Quantum Computing (RQC), Wako, Saitama 351-0198, Japan\\
$^{4}$Department of Physics, The University of Tokyo, 5-1-5 Kashiwanoha, Kashiwa, Chiba 277-8574, Japan\\
$^{5}$Quantum Information Research Center and Jiangxi Province Key Laboratory of Applied Optical Technology, Shangrao Normal University, Shangrao 334001, China\\
$^{6}$College of Sciences, Northeastern University, Shenyang 110819, China
}}

\date{\today}

\begin{abstract}
Quantum routers, as key components of quantum networks, are capable of coherent transmission of information carriers between quantum nodes at long distances in Markovian approximations and have received widespread attention in recent years, which have not been explored in non-Markovian systems. In this paper, we study a nonreciprocal quantum router with non-Markovian environments, which enables the directional control of single photons, permitting transmission from one side while preventing it from the opposite direction. The cascade system under study consists of two quantum nodes: one comprising two coupled coplanar-waveguide resonators and the other featuring a superconducting ring resonator. Each node is respectively coupled to a single Yttrium iron garnet (YIG) disk, with nonreciprocity arising from the selective coupling between magnons and microwave photons in our model. We analytically derive the transmission and reflection spectra of the system when a photon is input respectively from the left and right sides of the transmission line in the non-Markovian regimes. The results indicate that under suitable parameter conditions, a single-photon signal can be routed from a specific input port to either of the two output ports, while being fully absorbed when incident from the opposite port. Moreover, we compare the scattering behavior in the non-Markovian regimes with the results in the Markovian cases by numerical simulations, where the difference lies in that there exist two peaks with a peak value of unity in the transmission spectrum (two valleys with a minimum value of zero) under non-Markovian environments, while there is only a small range near detuning equal to zero to make the transmission spectrum up to a value of unity in the Markovian cases when the single photon inputs from the left side of the transmission line. As the environmental spectrum widths increase, the results given by the non-Markovian regimes are in good agreement with those in the Markovian cases. The formalism presented might open an alternative field of possible applications in quantum information and quantum communication with non-Markovianities.
\end{abstract}


\maketitle
\section{Introduction}
It is well known that quantum communications represent very important parts
of a rapidly developing research area called quantum information processing \cite{Nielsen2000,Bouwmeester2001,Saffman2313,Suter041001,Galindo347,Duan1209,Zhou023838}. Different from the classical information processing, quantum information processing is a brand-new way of information processing since the carrier of information in quantum information processing is quantum states \cite{Biolatti5647,Ahn012302,Zanardi4752} rather than the classical bits, leading it to own the unique advantages in achieving higher information processing efficiency and providing more storage space of information, which can also be robust \cite{Shi022441}. Within quantum information processing, the fundamental physical framework is provided by quantum networks \cite{Kimble1023,Lodahl347,Chang685,Zhao043703}, which are formed by interconnected quantum nodes and channels \cite{Cirac2331,Cirac1999,DiVincenzo771,Ritter195,Walmsley040001}. Moreover, in quantum information science one would like to distribute information over large distances. Photons are important candidates for such purposes because they are fast, readily available, and can travel over long distances without much decoherence, which are also required in quantum networks \cite{Kimble1023}. We need to route the photons as different nodes in the networks have to be linked. One of the most relevant devices for the operation of a quantum network is a quantum router--a quantum-mechanical counterpart of the classical router used to steer the information from its
source to intended destination, whose primary function in the simplest configuration is to send or route an incident photon into one of the two output channels. In contrast to a classical router, the quantum router \cite{Zueco042303,Pemberton-Ross020503,Hoi073601,Xia031013,Naaman112601,Pechal024009,
Wang014049} exploits quantum phenomena. In fact, a quantum router can coherently transfer the information carrier between distant quantum nodes, which is highly desirable and vital for practical applications.

The past decade has witnessed growing interest and rapid advancements in the field of quantum routing \cite{Lemr062333,Lu013805,Bartkiewicz022335,Korzeczek063714,Wanisch032624,Wang044050,
Christensen013004,Zhao033484,Zhang170568,Zhang1054299,Zhang094303,Zhou063703,
Kunzelmann032617,Bayat187204,Wu041030}. Various types of single photon quantum routers have also been proposed, including the ones based on the circuit quantum electrodynamics (QED) \cite{Papon524}, superconducting circuits \cite{Kemp104505,Liu067003,Niskanen723,Xiong032318,Yan609,Li055101}, waveguide-coupled resonators \cite{Yan4820,Zhu063815,Yang063809,Liu207}, cavity optomechanical platforms \cite{Weis1520,Jiang033113,Agarwal021801,Gao155503,Shang2023arXiv,Fang489,Ma015204,Ma035201,
Liu065501,Wang065202,Wang0652012023}, quantum dots \cite{Bentham221101}, atom systems \cite{Shen2331,Zhou103604,Shomroni903,Li8861,Wang40116,Li063836,Xia052315,Ahumada033827,
Poudyal043048,Yan023821}, Su-Schrieffer-Heeger model \cite{Qi023037,Zheng054037}, large detuning \cite{Wu054007}, cavity QED \cite{Aoki083601,Rosenblum033854,Krimer033820}, and so on \cite{Bao133,Du79,Li553,Lemr282,Cao32,Du24001,Chen583,Li39343,Lu22955,Lu13694,
Sala275701,Tian557,Wang035204,Yan64005,Yuan12452,Habraken115004,
Hu45582,Ko125605,Lee752022,Andriichuk91,Gao053011,Zhang053707,Palaiodimopoulos032622,
Wang063715,LiL061002}. Moreover, nonreciprocal routers \cite{Ren013711,Yan105102,Li023801,Wang015025,Wang065201,Yang369,Ren244002} have been extensively investigated in recent years. Nonreciprocal photonic transport allows the propagation of photons in one direction while blocking propagating in the opposite one. The optical reciprocity refers to the existence of Lorentz reciprocity in the system, where the amplitude remains the same \cite{Ranzani023024,Jalas579} when the light transmission direction is opposite, while optical nonreciprocity is to break this invariance. Devices with nonreciprocity properties can greatly simplify the construction of optical networks and have a wide range of applications in signal transmission. Therefore, the study of optical nonreciprocity is of great significance \cite{Peng033507,Kamal311,Potton717,Sun023520,Xiang043702,Jiao064008,Li053522,
Jiao143605,Xu044070,Jiang064037,Hu063516,Jing033707,Xu063845,Xu053853,Xu063808,Xu023827,Shang201908,Liu063701,Pan043505,Liu043716,Zheng063709,Qian043103}.

On the other hand, the dynamics of open quantum system is a long-standing
problem \cite{Breuer2002,Chen043853}, where all realistic quantum systems are open due to the unavoidable couplings to environments (of memory or memoryless) \cite{Li062124,ShangPRA2024,Franco1345053,ShangTQB2025,Caruso1203,Zhao619}. The Markovian approximations for open systems are only valid when the coupling between the system and the environment is weak and the characteristic time of the bath is sufficiently shorter than that of the system \cite{Breuer2002,Weiss2008,Li5614,Cui032209,Shen013826}, while non-Markovianity can be characterized by the information flow between the system and its environment \cite{Breuer021002,Breuer210401,Lorenzo020102,Shen033805042121}. The non-Markovian effect should be taken into account when the environment correlation time is longer than the characteristic time of the system, which occurs on many quantum systems including optical fields propagating in cavity arrays or in an optical fiber \cite{Hartmann849,Pellizzari5242,Biswas062303}, open quantum systems with Jaynes-Cummings models \cite{Shen032101012107,Shen062106,Shi105005}, trapped ions subjected to artificial colored noise \cite{Turchette053807,Myatt269,Maniscalco052101}, microcavities interacting with a coupled resonator optical waveguide or photonic crystals \cite{Vega015001,Groblacher7606,Leggett1987,Xu7389,Lin165330,Chang052105,Tan032102,Longhi063826,Bayindir2140,
Stefanou12127}, and so on \cite{Shen023856,Shen053705,Shen043714,Yang053712,Zhang033701,Li023712}.

However, we need to consider several questions: (i) Is it effective to generalize the quantum router from Markovian systems to non-Markovian ones? (ii) How can the non-Markovian effects influence the quantum routing? (iii) What are the differences and connections for routers between Markovian and non-Markovian environments?

To address these questions, we propose a scheme to achieve nonreciprocal quantum routing with non-Markovian environments through a hybrid waveguide-resonator system composed of two linearly coupled coplanar-waveguide (CPW) resonators and a superconducting ring resonator, where the former can be modeled as simple harmonic oscillators with two ground planes placed on either side of a narrow central conductor \cite{Blais032329}. A Yttrium iron garnet (YIG) disk is positioned above the superconducting ring resonator, where YIG is one of the widely adopted ferrimagnetic materials in magnomechanical systems due to its high spin density and low damping rate \cite{Zhang156401}. In particular, the ring resonator sustains two orthogonal chiral microwave modes, namely clockwise (CW) and counterclockwise (CCW) rotating modes \cite{Bliokh021803,Tang163901,Klingler072402,Xu053501,Ren053714}. Magnons and microwave photons interact exclusively when they share the same chirality \cite{Zhu043842,Yu064416,Zhang044039}; otherwise, their interaction is prohibited, leading to nonreciprocal behavior. In our hybrid waveguide-resonator system, a single photon entering from the left side of the transmission line remains decoupled from the magnon mode, allowing the two coupled CPW resonators to function as a quantum router, directing the photon to either of the two output ports. Conversely, when a single photon is introduced from the right side, it couples strongly to the magnon mode and experiences significant decay, resulting in full absorption. This implementation of nonreciprocal quantum routing can protect the signal resource from the influences of unexpected signals from the opposite direction. In addition, we investigate the scattering behaviors with non-Markovian environments by numerical simulations and compare the differences between the Markovian and non-Markovian cases. We find that when the spectrum width of non-Markovian environments is big enough, the transmission and reflection spectra are consistent with that under Markovian approximations.

The remainder of this paper is organized as follows. In Sec.~\ref{section II}, we describe the model setup of system. In Sec.~\ref{section III}, we study the quantum routing with non-Markovian environments. The motion equation of the system and non-Markovian input-output relations are given. We also derive the scattering matrix of the system and analyze the routing of incident photon when the photon is incident from both directions. In Sec.~\ref{section IV}, we discuss the scattering behaviors under Markovian approximations and compare it with that in the non-Markovian regimes. Finally, we conclude with a summary of the paper in Sec.~\ref{section V}.

\section{Model Setup}\label{section II}
\begin{figure}[t]
\centerline{
\includegraphics[width=7.6cm, height=3.0cm, clip]{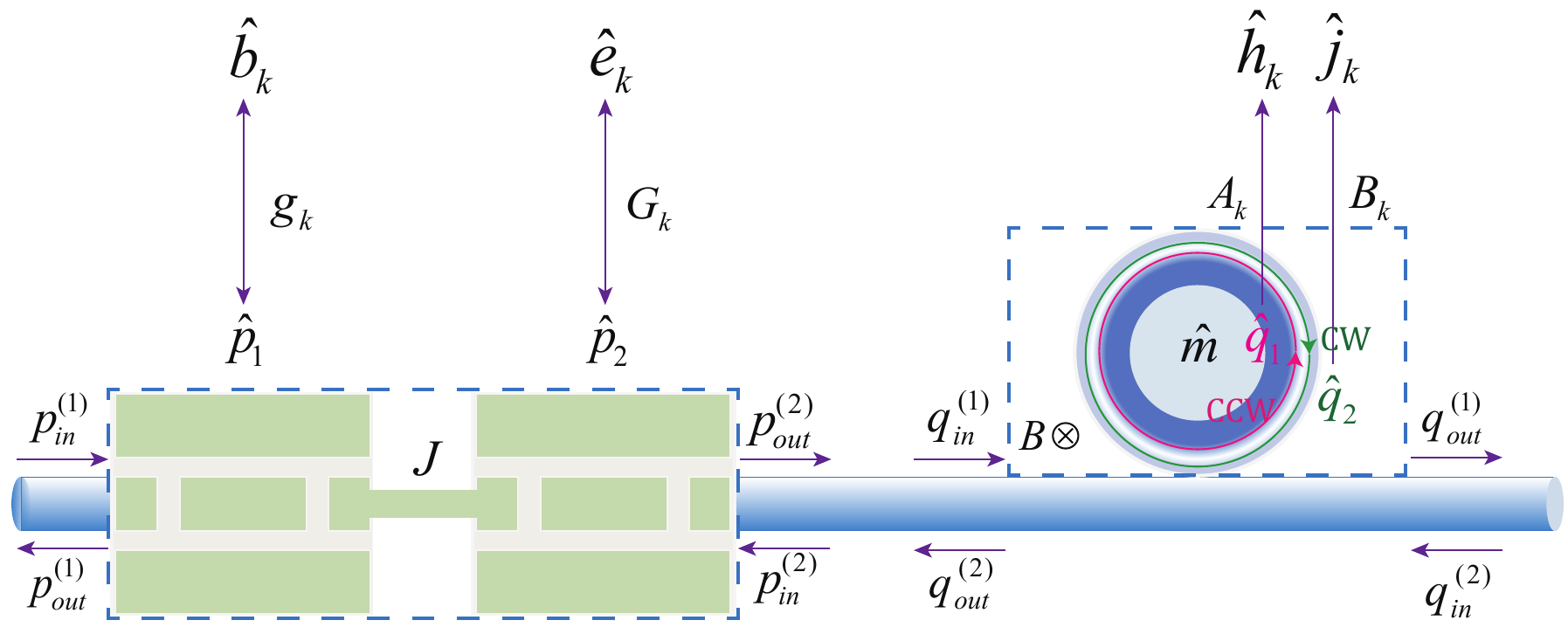}}
\caption{Schematic illustration of our system. The two dashed boxes represent two quantum nodes connected by a transmission line, where the first one denotes two coupled CPW resonators with annihilation operators $\hat p_1$ (interacting with non-Markovian environment by the coupling coefficient $g_k$) and $\hat p_2$ (interacting with non-Markovian environment by the coupling coefficient $G_k$), while the second one represents a superconducting ring resonator coupled to a single YIG disk. The superconducting ring resonator supports both CCW (The mode ${\hat q_1}$, interacts with a non-Markovian environment by the coupling coefficient $A_k$) and CW  (The mode ${\hat q_2}$, interacts with a non-Markovian environment by the coupling coefficient $B_k$) rotating microwave modes. The YIG disk with gray is positioned above the ring resonator and subjected to a perpendicular external static magnetic field $B$.} \label{fig1}
\end{figure}

The proposed model for the non-Markovian quantum router is shown in Fig.~\ref{fig1}, where the cascaded quantum system consists of two linearly CPW resonators and a superconducting ring resonator coupled to a single yttrium iron garnet (YIG) disk \cite{Ren013711,Zhang156401,Zhaoz1,Zhaoz2,Zhaoz3} (linked via a transmission line). In order to study the influences of the non-Markovian environments on the dynamics of the system, we take the coupling between the system and the environments into account. The total Hamiltonian containing non-Markovian environments can be written as
\begin{equation}
\begin{aligned}
\hat H = {\hat H_S} + {\hat H_E}
\label{H},
\end{aligned}
\end{equation}
with
\begin{align}
{\hat H_S} =& \hbar [{\omega _p}(\hat p_1^\dag {{\hat p}_1} + \hat p_2^\dag {{\hat p}_2}) + {\omega _q}(\hat q_1^\dag {{\hat q}_1} + \hat q_2^\dag {{\hat q}_2}) + {\omega _m}{{\hat m}^\dag }\hat m + \nonumber\\
&J(\hat p_1^\dag {{\hat p}_2} + \hat p_2^\dag {{\hat p}_1}) + \nu(\hat q_2^\dag \hat m + {{\hat m}^\dag }{{\hat q}_2})],
\label{H_S}
\end{align}
and
\begin{align}
{\hat H_E} = &{\rm{ }}\hbar [\sum\limits_k {{\omega _k}\hat b_k^\dag } {\hat b_k} + \sum\limits_k {({g_k}{{\hat p}_1}\hat b_k^\dag  + g_k^*\hat p_1^\dag {{\hat b}_k})}  + \sum\limits_k {{\Omega _k}} \hat e_k^\dag {\hat e_k} \nonumber\\
&+ \sum\limits_k {({G_k}{{\hat p}_2}\hat e_k^\dag  + G_k^*\hat p_2^\dag {{\hat e}_k})}  + \sum\limits_k {{\alpha _k}\hat h_k^\dag } {\hat h_k} \nonumber\\
&+ \sum\limits_k {({A_k}{{\hat q}_1}\hat h_k^\dag  + A_k^*\hat q_1^\dag {{\hat h}_k})}  + \sum\limits_k {{\beta _k}\hat j_k^\dag {{\hat j}_k}} \nonumber\\
&+ \sum\limits_k {({B_k}{{\hat q}_2}\hat j_k^\dag  + B_k^*\hat q_2^\dag {{\hat j}_k})}],
\label{expand_solution}
\end{align}
where $\hat p_1^\dag $ (${\hat p_1}$) and $\hat p_2^\dag $ (${\hat p_2}$) respectively denote the creation (annihilation) operator of two CPW resonators with identical frequency ${\omega _p}$. Here, $\hat q_1^\dag $ (${\hat q_1}$) represents the creation (annihilation) operator for the CCW mode of the superconducting ring resonator with the resonance frequency ${\omega _q}$, and $\hat q_2^\dag $ (${\hat q_2}$) describes the creation (annihilation) operator for the CW mode of the superconducting ring resonator with the resonance frequency ${\omega _q}$. ${\hat m^\dag }$ ($\hat m$) corresponds to the creation (annihilation) operator of the magnon mode with the resonance frequency ${\omega _m}$. The photon hopping rate is denoted by $J$. The resonators couple to the $k$th mode (eigenfrequencies ${\omega _k}$, ${\Omega _k}$, ${\alpha _k}$, and ${\beta _k}$) of the non-Markovian environments, which are modeled as collections of infinite modes via the creation (annihilation) operators $\hat b_k^\dag $ (${\hat b_k}$), $\hat e_k^\dag $ (${\hat e_k}$), $\hat h_k^\dag $ (${\hat h_k}$), and $\hat j_k^\dag $ (${\hat j_k}$), respectively. $\nu$ describes the coupling
strength between the CW microwave mode and the magnon mode. The parameters
$g_k$, $G_k$, $A_k$, and $B_k$ are coupling coefficients between four non-Markovian environments and resonators, respectively. It is worth mentioning that the terms related to $\hat q_1^\dag \hat m + {{\hat m}^\dag }\hat q_1 $ in Hamiltonian (\ref{H_S}) have been ignored because the magnon mode only couples to one of the propagating modes with the same chirality \cite{Zhang044039,Xie033701}. Magnons, being elementary excitations of spin waves, inherit the ability of single spins to precess, thereby also inherit chirality \cite{Sandratskii020406}. Therefore, their interaction with chiral microwave photons obeys a selection rule: only magnons and microwave photons with the same chirality interact with each other, while the interactions of magnons and photons with opposite chiralities are forbidden.
The decoupling of the magnon mode from the $\hat q_1$ mode breaks the time-reversal symmetry, thereby inducing nonreciprocal single-photon transmission.

\section{Quantum routing in non-Markovian environments} \label{section III}

\subsection{Equations of Motion and Non-Markovian Input-output Relations}
We now analytically derive the exact non-Markovian input-output relations. We shall use the equation of motion in the Heisenberg picture to solve the dynamics of the system subjected to the non-Markovian environments. The time evolutions of all annihilation operators including the CPW resonators, CCW and CW mode, magnon mode and environments are given by ${\hat p_1}(t) = {U^\dag }(t){\hat p_1}U(t)$, ${\hat p_2}(t) = {U^\dag }(t){\hat p_2}U(t)$, ${\hat q_1}(t) = {U^\dag }(t){\hat q_1}U(t)$, ${\hat q_2}(t) = {U^\dag }(t){\hat q_2}U(t)$, $\hat m(t) = {U^\dag }(t)\hat mU(t)$, ${\hat b_k}(t) = {U^\dag }(t){\hat b_k}U(t)$, ${\hat e_k}(t) = {U^\dag }(t){\hat e_k}U(t)$, ${\hat h_k}(t) = {U^\dag }(t){\hat h_k}U(t)$, and ${\hat j_k}(t) = {U^\dag }(t){\hat j_k}U(t)$, where $U(t) = \exp ( {{{ - i\hat Ht} \mathord{/
 {\vphantom {{ - i\hat Ht} \hbar }} 
 \kern-\nulldelimiterspace} \hbar }} )$ with $\hat H$ given by Eq.~(\ref{H}). According to the Heisenberg equations $\dot {\hat A}(t) =  - i[\hat A(t),\hat H(t)]/\hbar$ and the expectation values of the operators defined by ${p_1}(t) = \langle {{{\hat p}_1}}(t) \rangle $, ${p_2}(t) = \langle {{{\hat p}_2}}(t) \rangle $, ${q_1}(t) = \langle {{{\hat q}_1}}(t) \rangle $, ${q_2}(t) = \langle {{{\hat q}_2}}(t) \rangle $, $m(t) = \langle {\hat m}(t) \rangle $, ${b_k}(t) = \langle {{{\hat b}_k}}(t) \rangle $, ${e_k}(t) = \langle {{{\hat e}_k}}(t) \rangle $, ${h_k}(t) = \langle {{{\hat h}_k}}(t) \rangle $, and ${j_k}(t) = \langle {{{\hat j}_k}}(t) \rangle $, we can obtain
\begin{eqnarray}
\frac{d}{{dt}}{{p}_1}(t) &=&  - i{\omega _p}{{p}_1}(t) - iJ{{ p}_2}(t) - i\sum\limits_k {g_k^ * {{b}_k}(t)}, \label{c1}\\
\frac{d}{{dt}}{{p}_2}(t) &=&  - i{\omega _p}{{p}_2}(t) - iJ{{ p}_1}(t) - i\sum\limits_k {G_k^ * {{e}_k}(t)}, \label{c2}\\
\frac{d}{{dt}}{{q}_1}(t) &=&  - i{\omega _q}{{q}_1}(t) - i\sum\limits_k {A_k^ * } {{h}_k}(t),\label{d1}\\
\frac{d}{{dt}}{{q}_2}(t) &=&  - i{\omega _q}{{q}_2}(t) - i\nu m(t) - i\sum\limits_k {B_k^ * } {{j}_k}(t),\label{d2}\\
\frac{d}{{dt}}m(t) &=&  - i{\omega _m}m(t) - i\nu{{q}_2(t)}, \label{m}\\
\frac{d}{{dt}}{{b}_k}(t) &=&  - i\sum\limits_k {{\omega _k}{{b}_k}(t)}  - i\sum\limits_k {{g_k}{{p}_1}(t)}, \label{bk}\\
\frac{d}{{dt}}{{e}_k}(t) &=&  - i\sum\limits_k {{\Omega _k}{{e}_k}(t)}  - i\sum\limits_k {{G_k}{{p}_2}(t)}, \label{ek}\\
\frac{d}{{dt}}{{h}_k}(t) &=&  - i\sum\limits_k {{\alpha _k}{{h}_k}(t)}  - i\sum\limits_k {{A_k}{{q}_1}(t)}, \label{hk}\\
\frac{d}{{dt}}{{j}_k}(t) &=&  - i\sum\limits_k {{\beta _k}{{j}_k}(t)}  - i\sum\limits_k {{B_k}{{q}_2}(t)}. \label{jk}
\label{solution1}
\end{eqnarray}
Through simple calculations solving Eqs.~(\ref{bk})-(\ref{jk}), we
get the formal solutions of the environment operators for $t \ge 0$, i.e.,
\begin{equation}
\begin{aligned}
{{b}_k}(t) = {{b}_k}(0){e^{ - i{\omega _k}t}} - i{g_k}\int_0^t {d\tau {{p}_1}(\tau ){e^{ - i{\omega _k}(t - \tau )}}}, \\
{{e}_k}(t) = {{e}_k}(0){e^{ - i{\Omega _k}t}} - i{G_k}\int_0^t {d\tau {{p}_2}(\tau ){e^{ - i{\Omega _k}(t - \tau )}}}, \\
{{h}_k}(t) = {{h}_k}(0){e^{ - i{\alpha _k}t}} - i{A_k}\int_0^t {d\tau {{q}_1}(\tau ){e^{ - i{\alpha _k}(t - \tau )}}}, \\
{{j}_k}(t) = {{j}_k}(0){e^{ - i{\beta _k}t}} - i{B_k}\int_0^t {d\tau {{q}_2}(\tau ){e^{ - i{\beta _k}(t - \tau )}}}.
\label{env_solution1}
\end{aligned}
\end{equation}
The first terms on the right-hand sides of Eq.~(\ref{env_solution1}) denote the freely propagating parts of the environment fields, while the second terms describe the influences of the non-Markovian environments on the resonators. Substituting Eq.~(\ref{env_solution1}) into Eqs.~(\ref{c1})-(\ref{m}), we can obtain a set of integro-differential equations as
\begin{align}
\frac{d}{{dt}}{{p}_1}(t) = &- i{\omega _p}{{p}_1}(t) - iJ{{p}_2}(t) - K_1^*(t)  \nonumber\\
&- \int_0^t {d\tau {{p}_1}(\tau ){f_1}(t - \tau )}, \nonumber\\
\frac{d}{{dt}}{{p}_2}(t) = &- i{\omega _p}{{p}_2}(t) - iJ{{p}_1}(t) - K_2^*(t)  &\nonumber\\
&- \int_0^t {d\tau {{p}_2}(\tau ){f_2}(t - \tau )}, \nonumber\\
\frac{d}{{dt}}{{q}_1}(t)= &- i{\omega _q}{{q}_1}(t) - K_3^*(t)  - \int_0^t {d\tau {{q}_1}(\tau ){f_3}(t - \tau )},& \nonumber\\
\frac{d}{{dt}}{{q}_2}(t) = &- i{\omega _q}{{q}_2}(t) - i\nu m(t) - K_4^*(t) &\nonumber\\
& - \int_0^t {d\tau {{q}_2}(\tau ){f_4}(t - \tau )},\nonumber\\
\frac{d}{{dt}}m(t) =  &- i{\omega _m}m(t) - i\nu{{q}_2},
\label{Heisenberg_t>t0}
\end{align}
where the externally environments operators
\begin{align}
{K_1}(t) &= \int_{ - \infty }^\infty  {d\tau {h_1}(t - \tau )p_{in}^{(1)* }(\tau )}  \nonumber\\&=  - i\sum\limits_k {{e^{i{\omega _k}t}}{g_k}b_k^* (0)}, \nonumber\\
{K_2}(t) &= \int_{ - \infty }^\infty  {d\tau {h_2}(t - \tau )p_{in}^{(2)* }(\tau )}  \nonumber\\&=  - i\sum\limits_k {{e^{i{\Omega _k}t}}{G_k}e_k^* (0)}, \nonumber\\
{K_3}(t) &= \int_{ - \infty }^\infty  {d\tau {h_3}(t - \tau )q_{in}^{(1)* }(\tau )}  \nonumber\\&=  - i\sum\limits_k {{e^{i{\alpha _k}t}}{A_k}h_k^* (0)}, \nonumber\\
{K_4}(t) &= \int_{ - \infty }^\infty  {d\tau {h_4}(t - \tau )q_{in}^{(2)* }(\tau )}  \nonumber\\&=  - i\sum\limits_k {{e^{i{\beta _k}t}}{B_k}j_k^* (0)},
\label{K_1}
\end{align}
with
\begin{equation}
\begin{aligned}
p_{in}^{(1)}(t) &= \frac{1}{{\sqrt {2\pi } }}\sum\limits_k {{e^{ - i{\omega _k}t}}{{b}_k}(0)}, \\
p_{in}^{(2)}(t) &= \frac{{ - 1}}{{\sqrt {2\pi } }}\sum\limits_k {{e^{ - i{\Omega _k}t}}{{e}_k}(0)}, \\
q_{in}^{(1)}(t) &= \frac{1}{{\sqrt {2\pi } }}\sum\limits_k {{e^{ - i{\alpha _k}t}}{{h}_k}(0)}, \\
q_{in}^{(2)}(t) &= \frac{{ - 1}}{{\sqrt {2\pi } }}\sum\limits_k {{e^{ - i{\beta _k}t}}{{j}_k}(0)}.
\label{input}
\end{aligned}
\end{equation}
The impulse response functions in the continuum are
\begin{equation}
\begin{aligned}
{h_1}(t-\tau) &= \frac{{ - i}}{{\sqrt {2\pi } }}\int_{ - \infty }^\infty  {{e^{i\omega (t-\tau)}}g(\omega )d\omega }, \\
{h_2}(t-\tau) &= \frac{i}{{\sqrt {2\pi } }}\int_{ - \infty }^\infty  {{e^{i\omega (t-\tau)}}G(\omega )d\omega }, \\
{h_3}(t-\tau) &= \frac{{ - i}}{{\sqrt {2\pi } }}\int_{ - \infty }^\infty  {{e^{i\omega (t-\tau)}}A(\omega )d\omega }, \\
{h_4}(t-\tau) &= \frac{i}{{\sqrt {2\pi } }}\int_{ - \infty }^\infty  {{e^{i\omega (t-\tau)}}B(\omega )d\omega },
\label{k1234continuum}
\end{aligned}
\end{equation}
which define the impulse response functions that equal the Fourier
transform of the coupling strength $g(\omega )$, $G(\omega )$, $A(\omega )$, and $B(\omega )$ (here we have made the replacements ${g_k} \to g(\omega )$, ${G_k} \to G(\omega )$, ${A_k} \to A(\omega )$, and ${B_k} \to B(\omega )$).
The correlation functions in Eq.~(\ref{Heisenberg_t>t0}) are given by
\begin{align}
\!\!\!\!{f_n}(t){\rm{ }} \!=\! \int_{ - \infty }^{ + \infty } {{J_n}(\omega ){e^{ - i\omega t}}}  \!\equiv\!  \int_{ - \infty }^{ + \infty } {{h_n}( - \varepsilon ){h_n^*}(t - \varepsilon )} d\varepsilon,
\!\!\!\label{correlation}
\end{align}
where ${J_1}(\omega ) = {\left| {g(\omega )} \right|^2}$, ${J_2}(\omega ) = {\left| {G(\omega )} \right|^2}$, ${J_3}(\omega ) = {\left| {A(\omega )} \right|^2}$, and ${J_4}(\omega ) = {\left| {B(\omega )} \right|^2}$ denote the spectral densities of
the environments, while ${f_n}(t)$ $(n = 1,2,3,4)$ respectively denote the memory function of the four environments.
Similarly, when $t \le {t_1}$, by solving Eqs.~(\ref{bk})-(\ref{jk}), we obtain
\begin{equation}
\begin{aligned}
{{b}_k}(t) &= {{b}_k}({t_1}){e^{ - i{\omega _k}(t - {t_1})}} + i{g_k}\int_t^{{t_1}} {d\tau {{p}_1}(\tau ){e^{ - i{\omega _k}(t - \tau )}}}, \\
{{e}_k}(t) &= {{e}_k}({t_1}){e^{ - i{\Omega _k}(t - {t_1})}} + i{G_k}\int_t^{{t_1}} {d\tau {{p}_2}(\tau ){e^{ - i{\Omega _k}(t - \tau )}}}, \\
{{h}_k}(t) &= {{h}_k}({t_1}){e^{ - i{\alpha _k}(t - {t_1})}} + i{A_k}\int_t^{{t_1}} {d\tau {{q}_1}(\tau ){e^{ - i{\alpha _k}(t - \tau )}}}, \\
{{j}_k}(t) &= {{j}_k}({t_1}){e^{ - i{\beta _k}(t - {t_1})}} + i{B_k}\int_t^{{t_1}} {d\tau {{q}_2}(\tau ){e^{ - i{\beta _k}(t - \tau )}}}.
\label{env_solution2}
\end{aligned}
\end{equation}
Substituting Eq.~(\ref{env_solution2}) into Eqs.~(\ref{c1})-(\ref{m}), we can get the other set of integro-differential equations as
\begin{align}
\frac{d}{{dt}}{{p}_1}(t) =  &- i{\omega _p}{{p}_1}(t) - iJ{{p}_2}(t) - {{K'_1}(t)} &\nonumber\\
&+ \int_t^{{t_1}} {d\tau {{p}_1}(\tau ){f_1}(t - \tau )},& \nonumber\\
\frac{d}{{dt}}{{p}_2}(t) =  &- i{\omega _p}{{p}_2}(t) - iJ{{p}_1}(t) - {{K'_2}(t)} &\nonumber\\
&+ \int_t^{{t_1}} {d\tau {{p}_2}(\tau ){f_2}(t - \tau )},& \nonumber\\
\frac{d}{{dt}}{{q}_1}(t) =  &- i{\omega _q}{{q}_1}(t) - {{K'_3}(t)} + \int_t^{{t_1}} {d\tau {{q}_1}(\tau ){f_3}(t - \tau )},& \nonumber\\
\frac{d}{{dt}}{{q}_2}(t) =  &- i{\omega _q}{{q}_2}(t) - i\nu m(t) - {{K'_4}(t)} &\nonumber\\
&+ \int_t^{{t_1}} {d\tau {{q}_2}(\tau ){f_4}(t - \tau )}, \nonumber\\
\frac{d}{{dt}}m(t) = & - i{\omega _m}m(t) - i\nu{{q}_2(t)},&
\label{Heisenberg_t<t1}
\end{align}
where
\begin{small}
\begin{equation}
\begin{aligned}
{{K'_1}}(t) &= \int_{ - \infty }^\infty  {d\tau k_1^ * (t - \tau )p_{out}^{(1)}(\tau )}  = i\sum\limits_k {{e^{ - i{\omega _k}(t - {t_1})}}g_k^ * {{b}_k}({t_1})}, \\
{{K'_2}}(t) &= \int_{ - \infty }^\infty  {d\tau k_2^ * (t - \tau )p_{out}^{(2)}(\tau )}  = i\sum\limits_k {{e^{ - i{\Omega _k}(t - {t_1})}}G_k^ * {{e}_k}({t_1})}, \\
{{K'_3}}(t) &= \int_{ - \infty }^\infty  {d\tau k_3^ * (t - \tau )q_{out}^{(1)}(\tau )}  = i\sum\limits_k {{e^{ - i{\alpha _k}(t - {t_1})}}A_k^ * {{h}_k}({t_1})}, \\
{{K'_4}}(t) &= \int_{ - \infty }^\infty  {d\tau k_4^ * (t - \tau )q_{out}^{(2)}(\tau )}  = i\sum\limits_k {{e^{ - i{\beta _k}(t - {t_1})}}B_k^ * {{j}_k}({t_1})},
\label{K_1}
\end{aligned}
\end{equation}
\end{small}
with the defined output fields
\begin{equation}
\begin{aligned}
p_{out}^{(1)}(t) &= \frac{1}{{\sqrt {2\pi } }}\sum\limits_k {{e^{ - i{\omega _k}(t - {t_1})}}{{b}_k}({t_1})}, \\
p_{out}^{(2)}(t) &= \frac{{ - 1}}{{\sqrt {2\pi } }}\sum\limits_k {{e^{ - i{\Omega _k}(t - {t_1})}}{{e}_k}({t_1})}, \\
q_{out}^{(1)}(t) &= \frac{1}{{\sqrt {2\pi } }}\sum\limits_k {{e^{ - i{\alpha _k}(t - {t_1})}}{{h}_k}({t_1})}, \\
q_{out}^{(2)}(t) &= \frac{{ - 1}}{{\sqrt {2\pi } }}\sum\limits_k {{e^{ - i{\beta _k}(t - {t_1})}}{{j}_k}({t_1})}.
\label{output}
\end{aligned}
\end{equation}

Comparing Eq.~(\ref{Heisenberg_t>t0}) with Eq.~(\ref{Heisenberg_t<t1}) and using Eqs.~(\ref{input}) and (\ref{output}),  we can obtain the non-Markovian input-output relations (setting ${t_1} \to t$) as
\begin{equation}
\begin{aligned}
p_{out}^{(1)}(t) - p_{in}^{(1)}(t) &= \int_0^t {d\tau {{p}_1}(\tau ){h_1}(\tau  - t)}, \\
p_{out}^{(2)}(t) - p_{in}^{(2)}(t) &= \int_0^t {d\tau {{p}_2}(\tau ){h_2}(\tau  - t)}, \\
q_{out}^{(1)}(t) - q_{in}^{(1)}(t) &= \int_0^t {d\tau {{q}_1}(\tau ){h_3}(\tau  - t)}, \\
q_{out}^{(2)}(t) - q_{in}^{(2)}(t) &= \int_0^t {d\tau {{q}_2}(\tau ){h_4}(\tau  - t)},
\label{input_output_relation}
\end{aligned}
\end{equation}
where the impulse response functions in the continuum are given by Eq.~(\ref{k1234continuum}). In order to take the non-Markovian effects into account, the impulse response functions are set as
\begin{equation}
\begin{aligned}
{h_1}(t - \tau ) &=  - i\sqrt {{\Gamma _1}} {\lambda _1}{e^{{\lambda _1}(t - \tau )}}\Theta (\tau  - t),\\
{h_2}(t - \tau ) &= i\sqrt {{\Gamma _2}} {\lambda _2}{e^{{\lambda _2}(t - \tau )}}\Theta (\tau  - t),\\
{h_3}(t - \tau ) &=  - i\sqrt {{\Gamma _3}} {\lambda _3}{e^{{\lambda _3}(t - \tau )}}\Theta (\tau  - t),\\
{h_4}(t - \tau ) &= i\sqrt {{\Gamma _4}} {\lambda _4}{e^{{\lambda _4}(t - \tau )}}\Theta (\tau  - t),
\label{k1234theta}
\end{aligned}
\end{equation}
which leads to
\begin{equation}
\begin{aligned}
{f_n}(t - \tau ) = \frac{1}{2}{\Gamma _n}{\lambda _n}{e^{ - {\lambda _n}\left| {t - \tau } \right|}},
\label{fn}
\end{aligned}
\end{equation}
due to Eq.~(\ref{correlation}), where $\Theta (\tau  - t) = 1$ for $\tau  \ge t$ otherwise $\Theta (\tau  - t) = 0$. Making the Fourier transform to Eqs.~(\ref{k1234continuum}) and (\ref{correlation}), we can obtain the coupling strength and spectral density \cite{Breuer2002,Shen033835,Shen012156,Haikka052103}
\begin{equation}
\begin{aligned}
g(\omega ) = \sqrt {\frac{{{\Gamma _1}}}{{2\pi }}} \frac{{{\lambda _1}}}{{{\lambda _1} - i\omega }},\\
G(\omega ) = \sqrt {\frac{{{\Gamma _2}}}{{2\pi }}} \frac{{{\lambda _2}}}{{{\lambda _2} - i\omega }},\\
A(\omega ) = \sqrt {\frac{{{\Gamma _3}}}{{2\pi }}} \frac{{{\lambda _3}}}{{{\lambda _3} - i\omega }},\\
B(\omega ) = \sqrt {\frac{{{\Gamma _4}}}{{2\pi }}} \frac{{{\lambda _4}}}{{{\lambda _4} - i\omega }}.
\label{response}
\end{aligned}
\end{equation}
and
\begin{equation}
\begin{aligned}
{J_n}(\omega ) = \frac{{{\Gamma _n}}}{{2\pi }}\frac{{\lambda _n^2}}{{\lambda _n^2 + {\omega ^2}}},
\label{density}
\end{aligned}
\end{equation}
which represents the Lorentzian spectral density and is implemented by all-optical setups \cite{Xiong1000321012019,Cialdi1000521042019,Haseli900521182014,
Li1291405012022Xu820423282010Tang97100022012,Fanchini1122104022014} and pseudomode methods \cite{Jack630438032001,Barnett1997, Garraway5522901997,Garraway5546361997,Man900621042014,Man2357632015,
Mazzola800121042009,Pleasance960621052017,Tamascelli030402,Dalton053813,
Mazzola042302,Lazarou012331,Schonleber052108}. ${\lambda _n}$ $(n = 1,2,3,4)$ denotes the spectrum width of the non-Markovian environments, while  ${\Gamma _n}$ is the cavity decay rate through the input and output ports.

Now we consider the situation in the frequency domain. With the modified Laplace transform \cite{Shen052122} $\chi (\omega ) = \int_0^\infty  {dt{e^{ i\omega t}}\chi (t)} $, where ${e^{i\omega t}} \to {e^{i\omega t - \varepsilon t}}$ with $\varepsilon  \to {0^ + }$ makes $\chi (\omega )$ to converge to a finite value, we find that Eq.~(\ref{Heisenberg_t>t0}) in the frequency space can be rewritten as
\begin{widetext}
\begin{equation}
\begin{aligned}
&[i{\Delta _p} + {f_1}(\omega ) + {\gamma _p}]{{p}_1}(\omega ) + iJ{{p}_2}(\omega ) = {{\tilde k}_1}(\omega )p_{in}^{(1)}(i{\lambda _1}) - p_{in}^{(1)}(\omega ){{\tilde k}_1}(\omega ),\\
&[i{\Delta _p} + {f_2}(\omega ) + {\gamma _p}]{{p}_2}(\omega ) + iJ{{p}_1}(\omega ) = {{\tilde k}_2}(\omega )p_{in}^{(2)}(i{\lambda _2}) - p_{in}^{(2)}(\omega ){{\tilde k}_2}(\omega ),\\
&[i{\Delta _q} + {f_3}(\omega ) + {\gamma _q}]{{q}_1}(\omega ) = {{\tilde k}_3}(\omega )q_{in}^{(1)}(i{\lambda _3}) - q_{in}^{(1)}(\omega ){{\tilde k}_3}(\omega ),\\
&[i{\Delta _q} + {f_4}(\omega ) + {\gamma _q}]{{q}_2}(\omega ) + i\nu m(\omega ) = {{\tilde k}_4}(\omega )q_{in}^{(2)}(i{\lambda _4}) - q_{in}^{(2)}(\omega ){{\tilde k}_4}(\omega ),\\
&(i{\Delta _m} + {\gamma _m})m(\omega ) + i\nu{{q}_2}(\omega ) = 0,
\label{Heisenberg_fre}
\end{aligned}
\end{equation}
\end{widetext}
where ${{\tilde k}_n}(\omega ) = \int_{ - \infty }^0 {{e^{i\omega t}}k_1^ * (t)dt} $ $(n = 1,2,3,4)$. ${\Delta _p} = {\omega _p} - \omega $, ${\Delta _q} = {\omega _q} - \omega $, and ${\Delta _m} = {\omega _m} - \omega $ represent respectively the detunings. ${\gamma _p}$, ${\gamma _q}$, and ${\gamma _m}$ are the intrinsic damping rates of the superconducting CPW resonators, superconducting ring resonator and magnon mode, respectively.

In addition, according to Eq.~(\ref{input_output_relation}), we can obtain the non-Markovian input-output relations in the frequency domain as
\begin{equation}
\begin{aligned}
p_{out}^{(1)}(\omega ) - p_{in}^{(1)}(\omega ) = {{p}_1}(\omega ){h_1}( - \omega ),\\
p_{out}^{(2)}(\omega ) - p_{in}^{(2)}(\omega ) = {{p}_2}(\omega ){h_2}( - \omega ),\\
q_{out}^{(1)}(\omega ) - q_{in}^{(1)}(\omega ) = {{q}_1}(\omega ){h_3}( - \omega ),\\
q_{out}^{(2)}(\omega ) - q_{in}^{(2)}(\omega ) = {{q}_2}(\omega ){h_4}( - \omega ).
\label{input_output_relation_fre}
\end{aligned}
\end{equation}

\subsection{Nonreciprocal Quantum Router}
Based on the non-Markovian input-output relations (\ref{input_output_relation_fre}) together with $q_{in}^{(1)}(\omega ) = {e^{i\theta }}p_{out}^{(2)}(\omega )$ and $p_{in}^{(2)}(\omega ) = {e^{i\theta }}q_{out}^{(2)}(\omega )$, we can calculate the analytical expression of the output spectra, i.e., transmission and reflection spectra, where $\theta $ denotes the phase accumulated by the photon as it runs along the transmission line, expressed as the distance separating the CPW resonator and the ring resonator. According to Fig.~\ref{fig1}, the output fields can be easily written as ${M_{out}}(\omega ) = \mathcal {S}(\omega ){M_{in}}(\omega )$, where ${M_{in}}(\omega ) = [p_{in}^{(1)}(\omega ),q_{in}^{(2)}(\omega )]^T$ and ${M_{out}}(\omega ) = [q_{out}^{(1)}(\omega ),p_{out}^{(1)}(\omega )]^T$ are the vectors of the input and output
fields, respectively. The scattering matrix is defined as
\begin{equation}
\mathcal {S}(\omega ) = \left( {\begin{array}{*{20}{c}}
{{t_{pq}}}&{{r_{qq}}}\\
{{r_{pp}}}&{{t_{qp}}}
\end{array}} \right),
\label{scattering}
\end{equation}
with the relevant matrix elements
\begin{widetext}
\begin{equation}
\begin{aligned}
{t_{pq}}= &{e^{i\theta }}[1 - \frac{{{{\tilde k}_3}(\omega ){h_3}( - \omega )}}{{i{\Delta _q} + {f_3}(\omega ) + {\gamma _q}}} + \frac{{{{\tilde k}_3}(\omega )q_{in}^{(1)}(i{\lambda _3}){h_3}( - \omega )}}{{(i{\Delta _q} + {f_3}(\omega ) + {\gamma _q})q_{in}^{(1)}(\omega )}}] \{ \frac{{iJ{{\tilde k}_1}(\omega ){h_2}( - \omega )}}{{(i{\Delta _p} + {f_1}(\omega ) + {\gamma _p})(i{\Delta _p} + {f_2}(\omega ) + {\gamma _p}) + {J^2}}} &\\
&- \frac{{iJ{{\tilde k}_1}(\omega ){h_2}( - \omega )p_{in}^{(1)}(i{\lambda _1})}}{{[(i{\Delta _p} + {f_1}(\omega ) + {\gamma _p})(i{\Delta _p} + {f_2}(\omega ) + {\gamma _p}) + {J^2}]p_{in}^{(1)}(\omega )}}\}, \\
{r_{pp}}= &1 - \frac{{{{\tilde k}_1}(\omega )(i{\Delta _p} + {f_2}(\omega ) + {\gamma _p}){h_1}( - \omega )}}{{(i{\Delta _p} + {f_1}(\omega ) + {\gamma _p})(i{\Delta _p} + {f_2}(\omega ) + {\gamma _p}) + {J^2}}} + \frac{{{{\tilde k}_1}(\omega )(i{\Delta _p} + {f_2}(\omega ) + {\gamma _p}){h_1}( - \omega )p_{in}^{(1)}(i{\lambda _1})}}{{[(i{\Delta _p} + {f_1}(\omega ) + {\gamma _p})(i{\Delta _p} + {f_2}(\omega ) + {\gamma _p}) + {J^2}]p_{in}^{(1)}(\omega )}},\\
{t_{qp}}= &{e^{i\theta }}\{{\frac{{iJ{{\tilde k}_2}(\omega ){h_1}( - \omega )}}{{(i{\Delta _p} + {f_1}(\omega ) + {\gamma _p})(i{\Delta _p} + {f_2}(\omega ) + {\gamma _p}) + {J^2}}} - \frac{{iJ{{\tilde k}_2}(\omega )p_{in}^{(2)}(i{\lambda _2}){h_1}( - \omega )}}{{[(i{\Delta _p} + {f_1}(\omega ) + {\gamma _p})(i{\Delta _p} + {f_2}(\omega ) + {\gamma _p}) + {J^2}]p_{in}^{(2)}(\omega )}}}\} &\\
&\{ 1 + \frac{{(i{\Delta _m} + {\gamma _m}){{\tilde k}_4}(\omega )q_{in}^{(2)}(i{\lambda _4}){h_4}( - \omega )}}{{[(i{\Delta _q} + {f_4}(\omega ) + {\gamma _q})(i{\Delta _m} + {\gamma _m}) + {\nu^2}]q_{in}^{(2)}(\omega )}} - \frac{{(i{\Delta _m} + {\gamma _m}){{\tilde k}_4}(\omega ){h_4}( - \omega )}}{{[(i{\Delta _d} + {f_4}(\omega ) + {\gamma _q})(i{\Delta _m} + {\gamma _m}) + {\nu^2}]}}\}, \\
{r_{qq}}= &{e^{2i\theta }}[1 - \frac{{{{\tilde k}_3}(\omega ){h_3}( - \omega )}}{{i{\Delta _q} + {f_3}(\omega ) + {\gamma _q}}} + \frac{{{{\tilde k}_3}(\omega )q_{in}^{(1)}(i{\lambda _3}){h_3}( - \omega )}}{{(i{\Delta _q} + {f_3}(\omega ) + {\gamma _q})q_{in}^{(1)}(\omega )}}] [1 + \frac{{{{\tilde k}_2}(\omega )p_{in}^{(2)}(i{\lambda _2})(i{\Delta _p} + {f_2}(\omega ) + {\gamma _p}){h_2}( - \omega )}}{{[(i{\Delta _p} + {f_1}(\omega ) + {\gamma _p})(i{\Delta _p} + {f_2}(\omega ) + {\gamma _p}) + {J^2}]p_{in}^{(2)}(\omega)}} &\\
&- \frac{{{{\tilde k}_2}(\omega )(i{\Delta _p} + {f_2}(\omega ) + {\gamma _p}){h_2}( - \omega )}}{{(i{\Delta _p} + {f_1}(\omega ) + {\gamma _p})(i{\Delta _p} + {f_2}(\omega ) + {\gamma _p}) + {J^2}}}] \cdot \{ 1 + \frac{{(i{\Delta _m} + {\gamma _m}){{\tilde k}_4}(\omega )q_{in}^{(2)}(i{\lambda _4}){h_4}( - \omega )}}{{[(i{\Delta _q} + {f_4}(\omega ) + {\gamma _q})(i{\Delta _m} + {\gamma _m}) + {\nu^2}]q_{in}^{(2)}(\omega )}} &\\
&- \frac{{(i{\Delta _m} + {\gamma _m}){{\tilde k}_4}(\omega ){h_4}( - \omega )}}{{[(i{\Delta _q} + {f_4}(\omega ) + {\gamma _q})(i{\Delta _m} + {\gamma _m}) + {\nu^2}]}}\},
\label{elements}
\end{aligned}
\end{equation}
\end{widetext}
which determine the scattering behavior of an input single-photon state. Obviously, the scattering matrix (\ref{scattering}) is asymmetric, which arises from the unidirectional magnon-photon coupling, leading us to anticipate that the system will exhibit distinct phenomena depending on the input direction of the single photon. That is to say, when the single photon is input from the left and right side of the transmission line, the hybrid system will generate different physical responses, which can be reflected in the transmission and reflection spectra.

\begin{figure}[t!]
\centerline{
\includegraphics[width=8.2cm, height=6.5cm, clip]{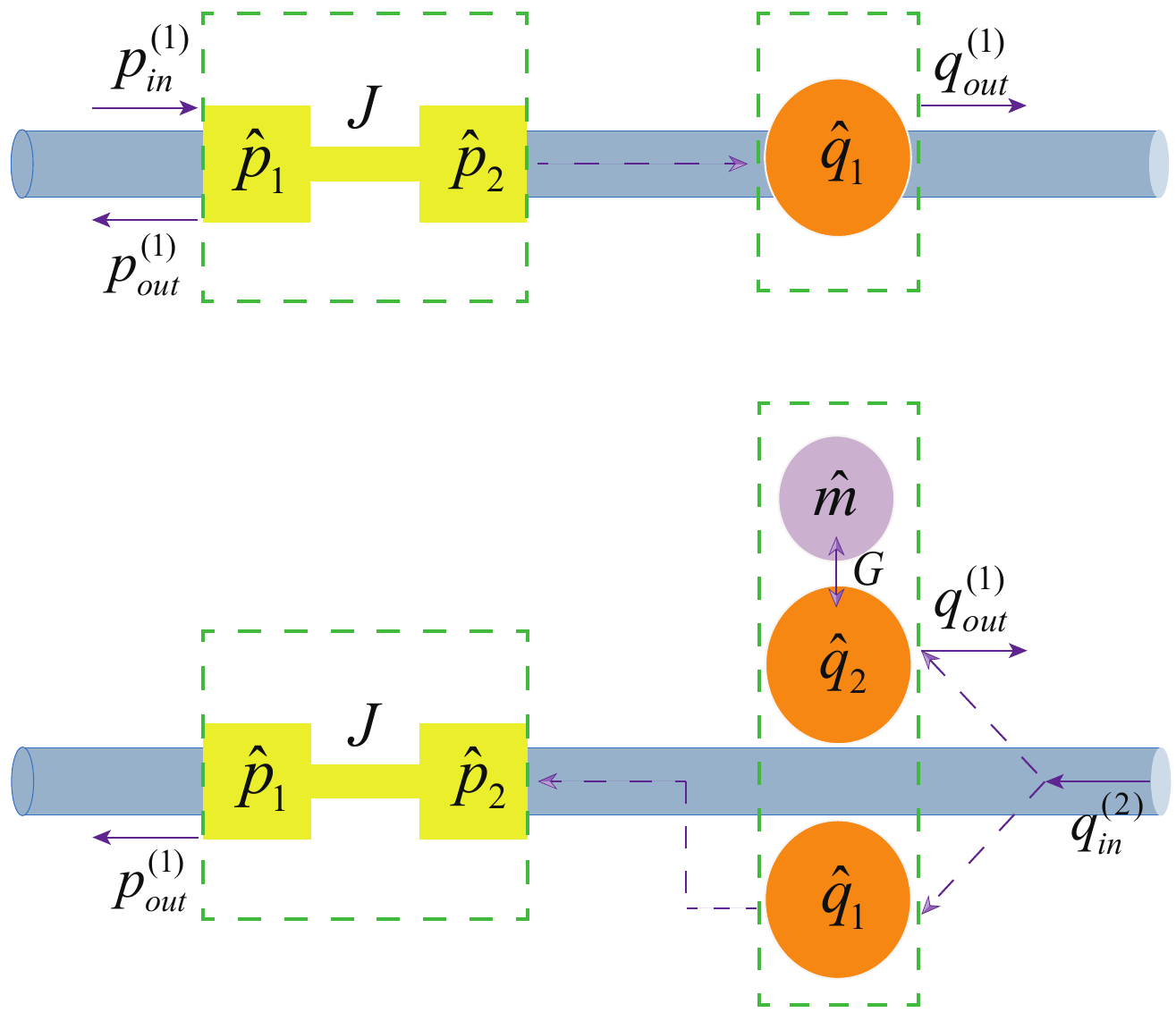}}
\caption{Schematic diagram of the nonreciprocal quantum router can be realized in a hybrid system consisting of two directly coupled coplanar-waveguide resonators and a superconducting ring resonator connected together through a transmission line. Nonreciprocal behavior here refers to the different physical phenomena occurring when the incident light propagates in different directions, which arises from broken time-reversal symmetry. (a) indicates the schematic diagram of the left-incident photon, while (b) shows the schematic diagram when the single photon is incident from the right side of the transmission line.}\label{fig2}
\end{figure}

Next, we will explore in detail the output spectra of the system when a single photon is incident from different directions. For convenience, we hereafter set ${\Delta _p} = {\Delta _q} = {\Delta _m} \equiv  \Delta $ and ${\Gamma _1} = {\Gamma _2} = {\Gamma _3} = {\Gamma _4} \equiv \Gamma$. We discuss the transmission and reflection spectra under non-Markovian regimes by considering the following two aspects.

\begin{figure}[t]
\centerline{
\includegraphics[width=7.0cm, height=11.0cm, clip]{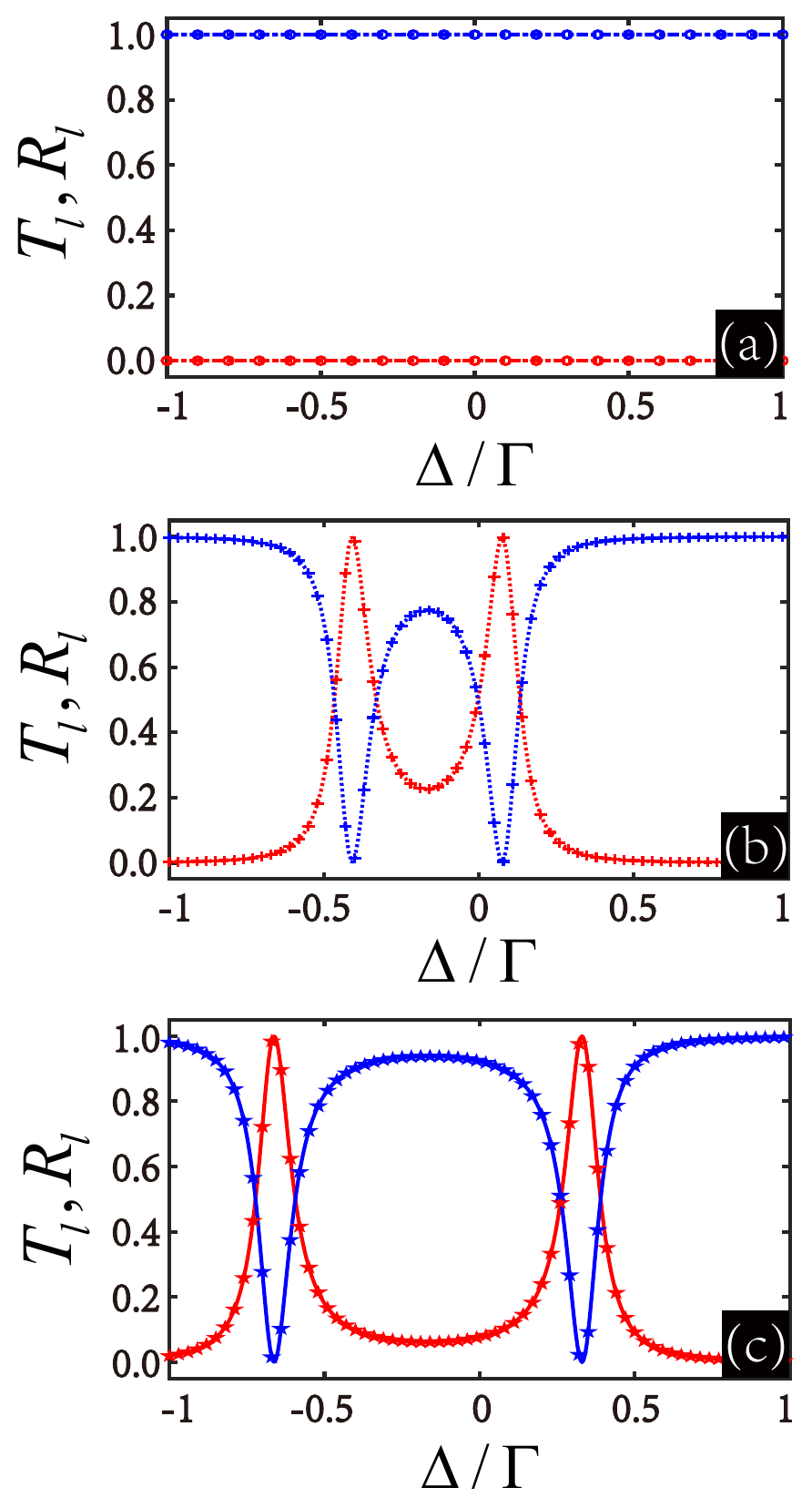}}
\caption{Transmission and reflection spectra in the case of the left-incident photon. (a)(b)(c) show the transmission (red lines) and reflection (blue lines) spectra ${T_l} = {\left| {{t_{pq}}} \right|^2}$ and ${R_l} = {\left| {{r_{pp}}} \right|^2}$ with ${t_{pq}}$ and ${r_{pp}}$ given by Eq.~(\ref{elements}) as a function of the detuning $\Delta /\Gamma $ for different photon hopping rate $J$, where $J=0$ (dotted-dashed lines) for (a), $J=0.25\Gamma$ (dotted lines) for (b), and $J=0.5\Gamma$ (solid lines) for (c), respectively. For convenience, we set ${\Delta _p} = {\Delta _q} \equiv  \Delta $ and ${\Gamma _1} = {\Gamma _2} = {\Gamma _3} = {\Gamma _4} \equiv \Gamma$. The parameters chosen are $\omega=2.5 \Gamma$, ${\gamma _p} = {\gamma _q}=0$, ${a_1} = 1,\gamma  = 0.001,{b_1} = 0.003$, $\theta=2 \pi$, ${\lambda _1} = {\lambda _2} =0.95\Gamma$, and $ {\lambda _3} = {\lambda _4} = \Gamma $.}\label{fig3}
\end{figure}

(i) In the first case, we assume the incident photon is injected from the left side of the transmission line, which means $q_{in}^{(2)} = 0$ and $p_{in}^{(1)} \ne 0$, as shown in Fig.~\ref{fig2}(a). In this incident direction, the superconducting ring resonator is decoupled from the magnon mode as mentioned earlier, so we have omitted this part of the structure in the schematic diagram. In other words, in this scenario, the two coupled CPW resonators primarily function as quantum routers and govern the scattering behavior of the left-input photon, where the output fields respectively yield $q_{out}^{(1)} = {t_{pq}}p_{in}^{(1)}$ and $p_{out}^{(1)} = {r_{pp}}p_{in}^{(1)}$ with ${t_{pq}}$, ${r_{pp}}$ given by Eq.~(\ref{elements}). We plot the transmission and reflection spectra ${T_l} = {\left| {{t_{pq}}} \right|^2}$ and ${R_l} = {\left| {{r_{pp}}} \right|^2}$ as functions of the detuning $\Delta /\Gamma $ under different hopping rate $J$ in Fig.~\ref{fig3}(a), where all red lines denote the transmission spectrum ${T_l}$ and blue lines show the reflection spectrum ${R_l}$, while the dotted-dashed, dotted, and solid lines represent $J = 0,0.25\Gamma,0.5\Gamma$, respectively. The circle, star, and plus lines respectively represent transmission ${T_l}$ and reflection ${R_l}$ in the case of ignoring the inhomogeneous terms ${\varphi _1} = {t_1}{\phi _1}({\lambda _3},\omega )$ and ${\varphi _2} = {t_2}{\phi _2}({\lambda _1},\omega )$ when $J = 0,0.25\Gamma,0.5\Gamma$, where the definitions of ${t _{1,2}}$, ${\phi _{1,2}}$ and related discussions on inhomogeneous terms are detailed in Appendix. For the input fields with the form of damped oscillation and given parameters, e.g., $\gamma=0.001$, ${b_1}=0.003$, we can calculate $\left| {{\phi _1}({\lambda _3},\omega )} \right|\sim {10^{ - 5}}$, $\left| {{\phi _2}({\lambda _1},\omega )} \right| \sim {10^{ - 5}}$ for $\lambda_1=0.95\Gamma$ and $\lambda_3=\Gamma$. This indicates that the influences of inhomogeneous terms $\left| {{\phi _1}} \right|$ and $\left| {{\phi _2}} \right|$ are small enough to be ignored, as demonstrated in Fig.~\ref{fig3}(a)(b)(c).

\begin{figure}[t]
\centerline{
\includegraphics[width=8.5cm, height=3cm, clip]{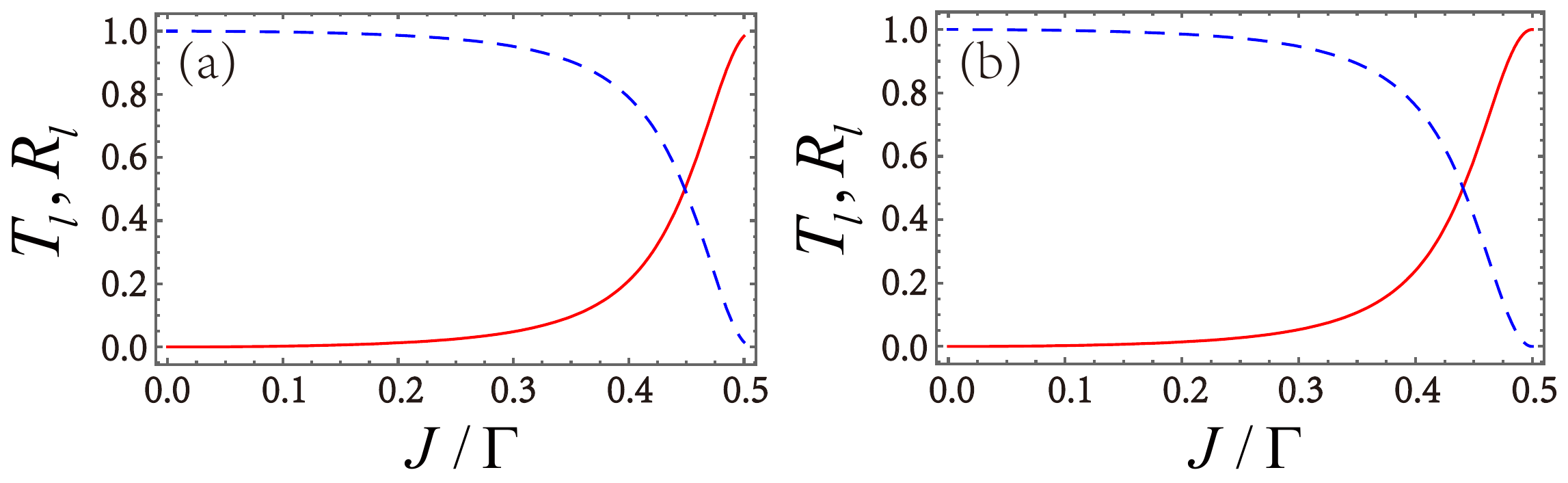}}
\caption{Transmission spectrum ${T_l}$ (red solid lines) and reflection spectrum ${R_l}$ (blue dashed lines) as a function of the photon hopping rate $J/\Gamma $ in the case of the left-incident photon. For convenience, we set ${\Delta _p} = {\Delta _q} \equiv \Delta $ and ${\Gamma _1} = {\Gamma _2} = {\Gamma _3} = {\Gamma _4} \equiv \Gamma$. The parameters chosen are $\omega=2.5 \Gamma$, ${\gamma _p} = {\gamma _q}=0$, ${a_1} = 1,\gamma  = 0.001$, ${b_1} = 0.003$, $\theta=2 \pi$, ${\lambda _1} = {\lambda _2} =0.95\Gamma$, and $ {\lambda _3} = {\lambda _4} = \Gamma $, where $\Delta=-0.67\Gamma$ for (a) while $\Delta=0.33\Gamma$ for (b).}\label{TlRlJ}
\end{figure}

In addition, from Fig.~\ref{fig3}, we can see that the transmission and reflection spectra are influenced by the hopping rate $J$, where the transmission coefficient ${T_l}=0$ and reflection coefficient ${R_l}=1$ for $J = 0$ (see Fig.~\ref{fig3}(a)), while the coefficients ${T_l}$ and ${R_l}$ are respectively close to $1$ and $0$ obtained for $J = 0.5\Gamma$ around the central resonance frequency $\Delta = -0.67\Gamma$ and $\Delta = 0.33\Gamma$ (see Fig.~\ref{fig3}(c)). Similar to Fig.~\ref{fig3}(c), Fig.~\ref{fig3}(b) shows that the transmission and reflection can also reach close to $1$ and $0$ at appropriate values of $\Delta$ when $J = 0.25\Gamma$, but the corresponding detuning positions are different. This also reflects that the hopping rate $J$ has influences on the scattering spectra. Therefore, we plot the transmission and reflection probabilities as a function of the photon hopping rate $J$ in Fig.~\ref{TlRlJ}, with the detuning $\Delta = -0.67\Gamma$ for Fig.~\ref{TlRlJ}(a) and $\Delta = 0.33\Gamma$ for Fig.~\ref{TlRlJ}(b). The results show that tuning the hopping rate $J$ from $0$ to $0.5\Gamma$ allows the transmission (reflection) probability to be smoothly modulated from $0 (1)$ to $1 (0)$ for both detuning positions, facilitating efficient routing of the left-input photon. This also indicates that the transport of left-input photon can be adjusted as needed by tuning the hopping rate $J$ of the two CPW resonators.

\begin{figure}[t]
\centerline{
\includegraphics[width=8.9cm, height=4cm, clip]{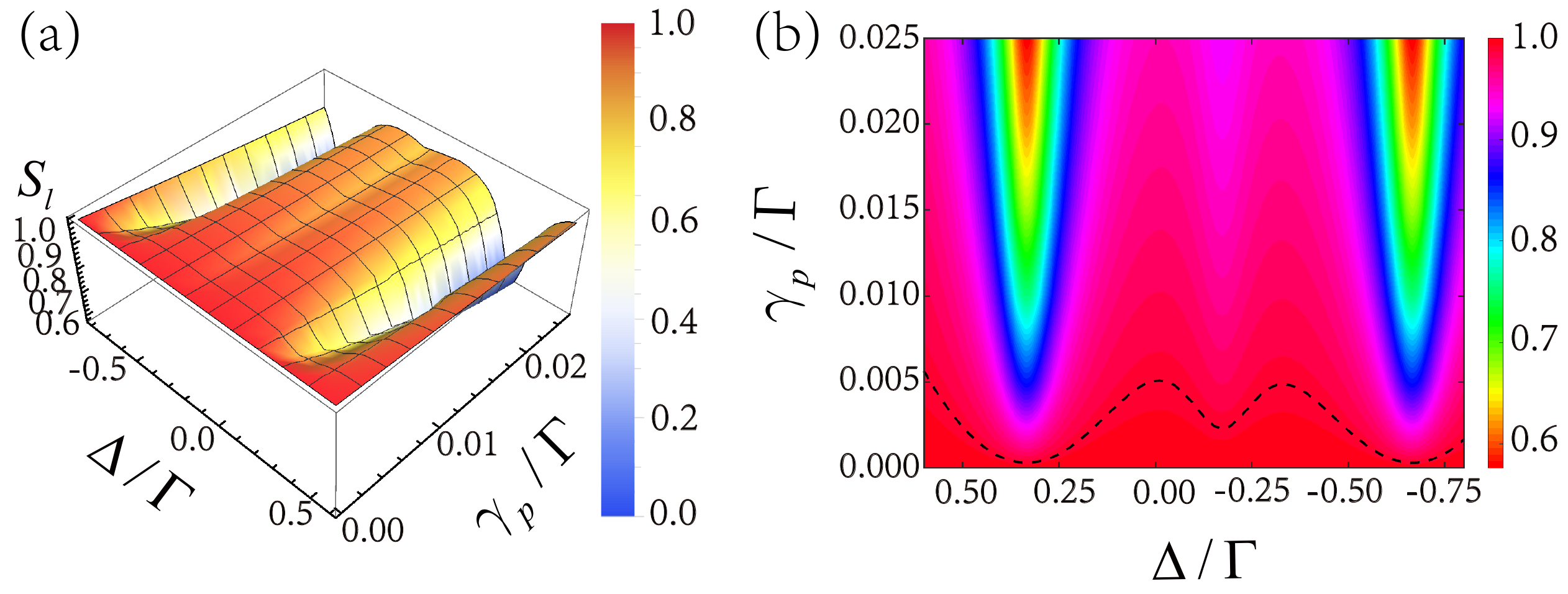}}
\caption{(a) Total scattering spectrum ${S_l}$ versus the damping rate $\gamma_p /\Gamma $ and detuning $\Delta /\Gamma $ for the left-input photon with ${S_l} = {T_l} + {R_l}$. The parameters are chosen as $\omega  = 2.5\Gamma$ and $J=0.5\Gamma$. For convenience, we set ${\Delta _p} = {\Delta _q} \equiv \Delta $ and $\gamma_q=\gamma_p$. (b) The total scattering spectrum ${S_l}$ as a function of the damping rate $\gamma_p /\Gamma $ and detuning $\Delta /\Gamma $ when the single photon is input from the left side of the transmission line is shown by the pseudocolor. The black dashed line represents $S_l=0.99$.}\label{fig4}
\end{figure}

Figure \ref{fig4} plots the total spectrum ${S_l}={T_l}+{R_l}$ as a function of the detuning $\Delta /\Gamma $ and damping rate $\gamma_p /\Gamma $ when the photon is input from the left side of the transmission line. In practical applications, the inherent photon losses of microwave resonators inevitably affect their routing performance. Indeed, we can see from Fig.~\ref{fig4} that the total spectrum ${S_l}$ diminishes as the damping rate increases, but it consistently exceeds 0.6 within the parameter range we selected. This also indicates that the routing capability of the non-Markovian router we studied is still sufficient under a relatively small intrinsic damping rate $\gamma_p$.

From the perspective of experimental implementation, the efficient scattering results ($S_l>0.6$) we obtained are realized at the selected parameters $\gamma_p /\Gamma =[0,0.025]$. If we take $\Gamma/2\pi=40$ MHz, the intrinsic decay rate $\gamma_p/2\pi$ should be less than 1 MHz to achieve the efficient results shown in Fig.~\ref{fig4}(a), which causes the needed internal quality factor $Q=\omega_0/\gamma_p>6\times10^3$ for a cavity frequency $\omega_0/2\pi=6$ GHz. It is easy to achieve the cavity quality factor of such a magnitude experimentally. The almost perfect results can even be achieved, namely $S_l>0.99$, as shown in Fig.~\ref{fig4}. Under the same parameter values $\Gamma/2\pi=40$ MHz, the intrinsic decay rate $\gamma_p/2\pi$ needs to less than 0.005 MHz and the internal quality factor $Q=\omega_0/\gamma_p>3\times10^4$ is required to reach $S_l>0.99$. Experimental results \cite{Megrant113510,Romanenk034032} have shown that superconducting cavities can achieve internal quality factors $Q>10^7$, confirming the realizability of our quantum routing scheme with existing advancements.

\begin{figure}[t]
\centerline{
\includegraphics[width=8.5cm, height=3cm, clip]{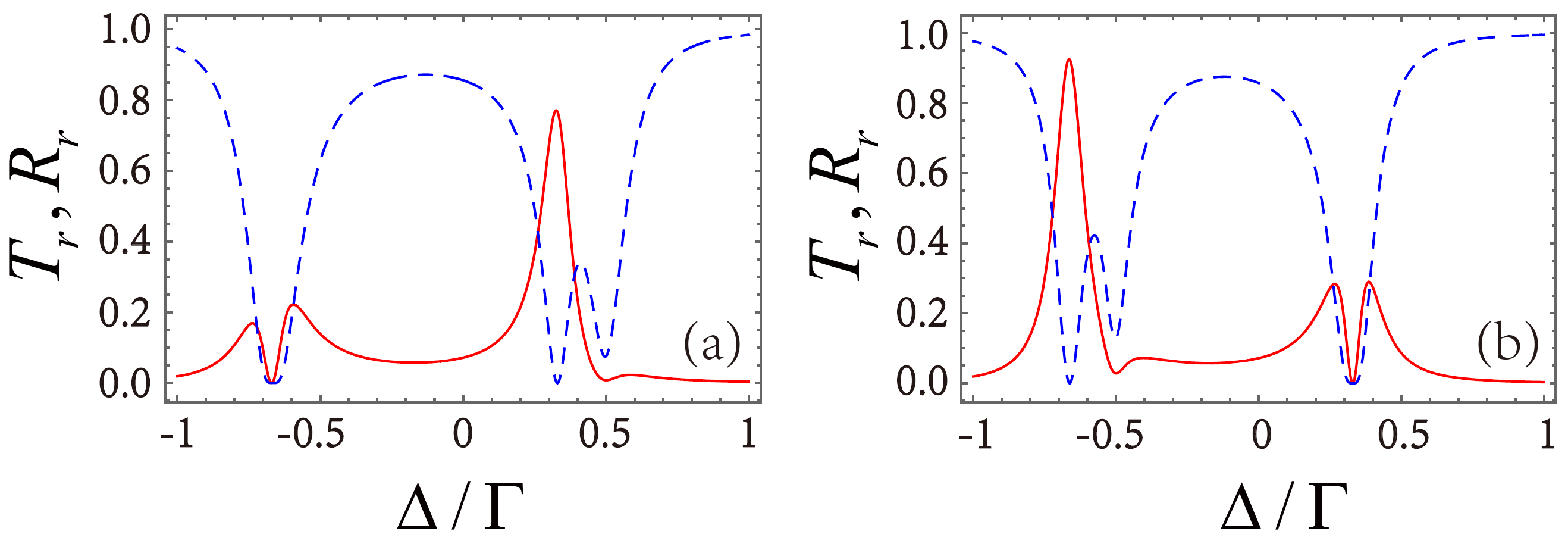}}
\caption{Transmission spectrum ${T_r}$ and reflection spectrum ${R_r}$ as a function of the detuning $\Delta /\Gamma $ when the photon is incident from the right side of the transmission line. The red solid and blue dashed lines respectively represent the transmission spectrum ${T_r} = {\left| {{t_{qp}}} \right|^2}$ and reflection spectrum ${R_r} = {\left| {{r_{qq}}} \right|^2}$, where ${t_{qp}}$ and ${r_{qq}}$ are given by Eq.~(\ref{elements}). For convenience, we set ${\Delta _p} = {\Delta _q} = {\Delta _m} \equiv \Delta  $ and ${\Gamma _1} = {\Gamma _2} = {\Gamma _3} = {\Gamma _4} \equiv \Gamma $. The parameters chosen are ${\gamma _m} \approx 0.093 \Gamma$, $\nu \approx 0.583 \Gamma$ for (a) but ${\gamma _m} \approx 0.045 \Gamma$, $\nu \approx 0.411 \Gamma$ for (b). The other parameters are ${a_1} = 1,\gamma  = 0.001$, ${b_1} = 0.003$, $\theta=2 \pi$, ${\gamma _p} = {\gamma _q}=0$, $J=0.5 \Gamma$, ${\lambda _1} = {\lambda _2} = 0.95 \Gamma$, and ${\lambda _3} = {\lambda _4} = \Gamma $.}\label{fig5}
\end{figure}

\begin{figure}[t]
\centerline{
\includegraphics[width=8.6cm, height=3.6cm, clip]{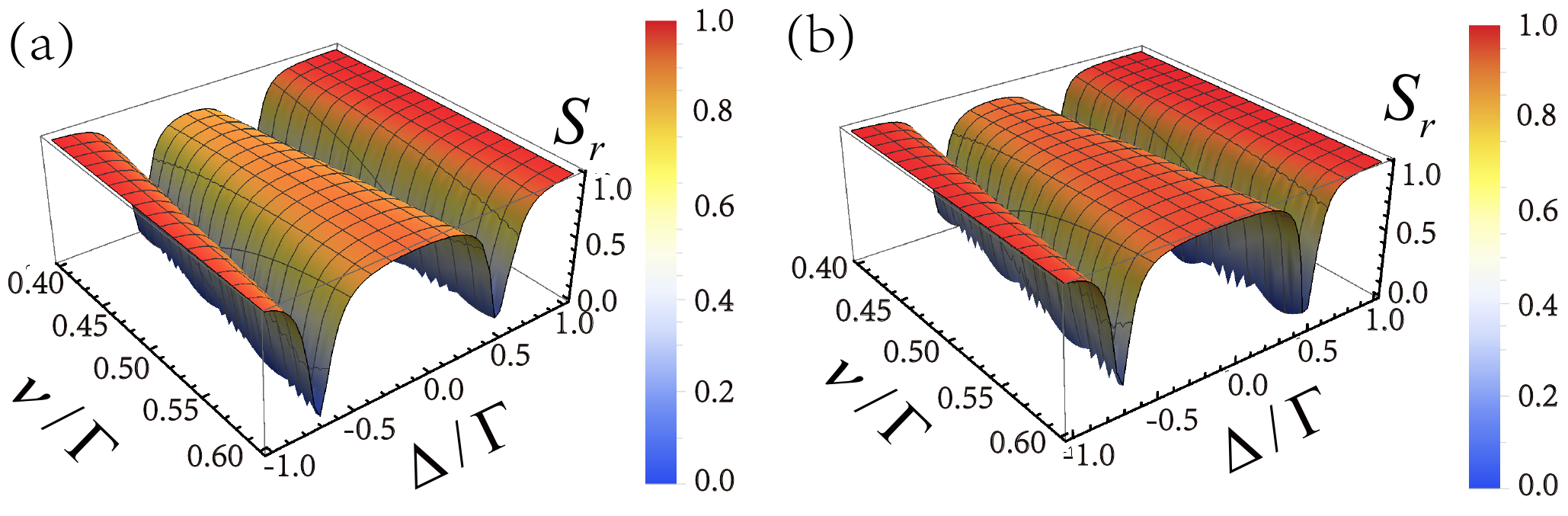}}
\caption{Total scattering spectrum ${S_r} = {T_r} + {R_r}$ versus the coupling strength $\nu /\Gamma $ and detuning $\Delta /\Gamma $ in the case of the right-incident photon. The parameters are chosen as ${\gamma _m} \approx 0.093 \Gamma$ for (a) and ${\gamma _m} \approx 0.045 \Gamma$ for (b), $\nu/\Gamma  = [0.4,0.6]$ and $\Delta /\Gamma  = [ - 1,1]$. Other parameters are the same as those in Fig.~\ref{fig5}.}\label{fig6}
\end{figure}

(ii) In the second case, we consider the case of the right-input photon, that is to say, $p_{in}^{(1)} = 0$ and $q_{in}^{(2)} \ne 0$ as shown in Fig.~\ref{fig2}(b). At this time, the magnon mode $\hat m$ couples with clockwise mode ${\hat q_2}$ with the same chirality and the output fields yield $p_{out}^{(1)} = {t_{qp}}q_{in}^{(2)}$ and $q_{out}^{(1)} = {r_{qq}}q_{in}^{(2)}$, where ${t_{qp}}$ and ${r_{qq}}$ are given by Eq.~(\ref{elements}). The transmission and reflection probabilities of the right-going photon are denoted by ${T_r} = {\left| {{t_{qp}}} \right|^2}$ and ${R_r} = {\left| {{r_{qq}}} \right|^2}$, respectively. We plot the transmission spectrum ${T_r}$ and reflection spectrum ${R_r}$ as a function of the detuning $\Delta /\Gamma $ in Fig.~\ref{fig5}. From Fig.~\ref{fig5}(a) we can see that both ${T_r}$ and ${R_r}$ are close to $0$ at the point $\Delta /\Gamma =-0.67$ under selected parameters, which suggests that the right-going photon is completely absorbed. The parameters we select here satisfy ${T_r}=0$ and ${R_r}=0$ with the expression of ${T_r}$ and ${R_r}$ given by Eq.~(\ref{elements}). That is to say, the total absorption of the right-going photon can be achieved under control of more than one set of parameters. To verify this, we plot Fig.~\ref{fig5}(b) under different parameters, from which we find that the transmission and reflection spectra
${T_r}=0$ and ${R_r}=0$ when $\Delta /\Gamma =0.33$. This also indicates that the total absorption of our quantum router when the photon input from the right side is fully tunable. In addition, we also plot the total scattering spectrum ${S_r}={T_r}+{R_r}$ as a function of the detuning $\Delta /\Gamma $ and coupling strength $\nu /\Gamma $ in Fig.~\ref{fig6} in the case of the right-incident photon. We show that the total spectrum $S_r$ can be zero under appropriate parameters, which refers to the input photon is completely absorbed.

\begin{figure}[t]
\centerline{
\includegraphics[width=8.5cm, height=3.8cm, clip]{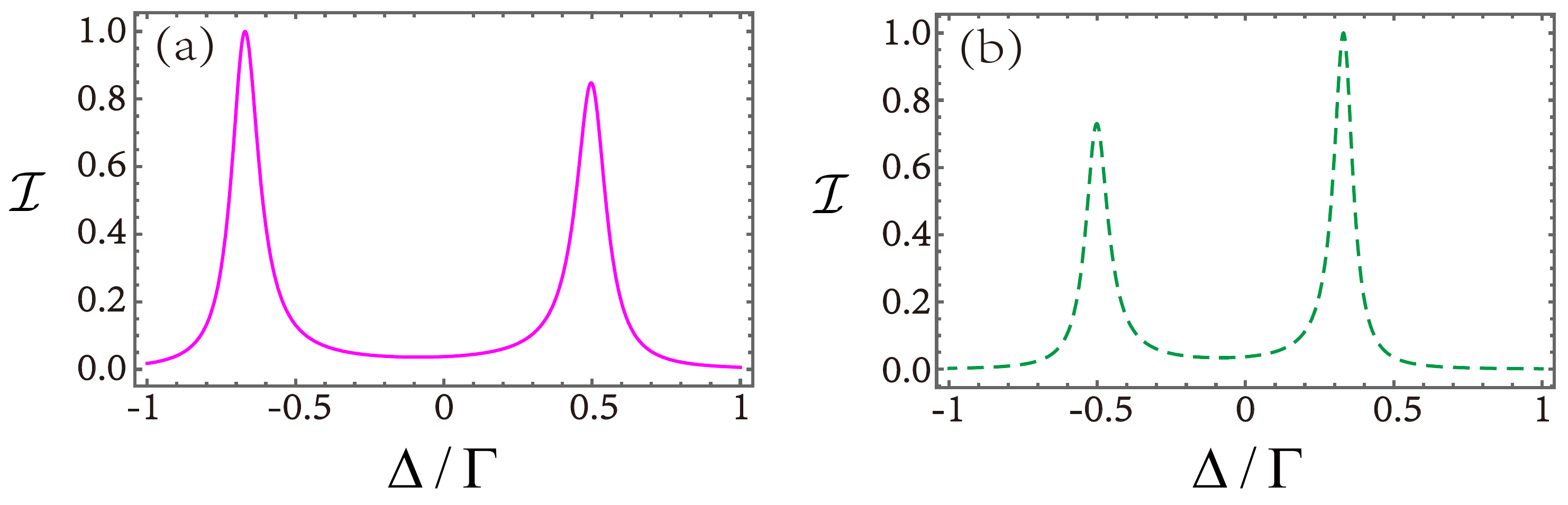}}
\caption{Contrast ratio $\mathcal {I}$ given by Eq.~(\ref{ratio}) versus the detuning $\Delta /\Gamma $. The parameters are chosen as ${\gamma _m} \approx 0.093 \Gamma$, $\nu \approx 0.583 \Gamma$ for (a), and ${\gamma _m} \approx 0.045 \Gamma$, $\nu \approx 0.411 \Gamma$ for (b). The other parameters are the same as those in Fig.~\ref{fig5}. }\label{fig7}
\end{figure}

Based on the analysis of the above two parts (i) and (ii), we show that an input photon from the left side of the transmission line can be routed to one of the two output channels, while an input photon from the right side
of the transmission line can be completely absorbed by the dissipative magnon mode, which reflects the nonreciprocal characteristics. For a deeper understanding of the quantum router's nonreciprocity, we evaluate the contrast ratio $\mathcal {I}$ given by
\begin{equation}
\begin{aligned}
\mathcal {I}  = \left| {\frac{{{S_l} - {S_r}}}{{{S_l} + {S_r}}}} \right|.
\label{ratio}
\end{aligned}
\end{equation}
Obviously, when a single photon is input from the left with full transmission and input from the right with full absorption, the contrast ratio can reach $1$. We show the contrast ratio $\mathcal {I}$ as a function of the detuning $\Delta /\Gamma $ in Fig.~\ref{fig7}. From Fig.~\ref{fig7}(a), we find that $\mathcal {I}$ is close to $1$ when $\Delta /\Gamma =-0.67$, which indicates the nonreciprocal single-photon router can be performed as an one-way quantum router. Similarly, Fig.~\ref{fig7}(b) shows that there exists another position for the detuning $\Delta /\Gamma =0.33$ to make the contrast ratio $\mathcal {I}$ almost 1.

\section{Discussions on the Markovian regimes} \label{section IV}
\begin{figure}[t]
\centerline{
\includegraphics[width=8cm, height=8.8cm, clip]{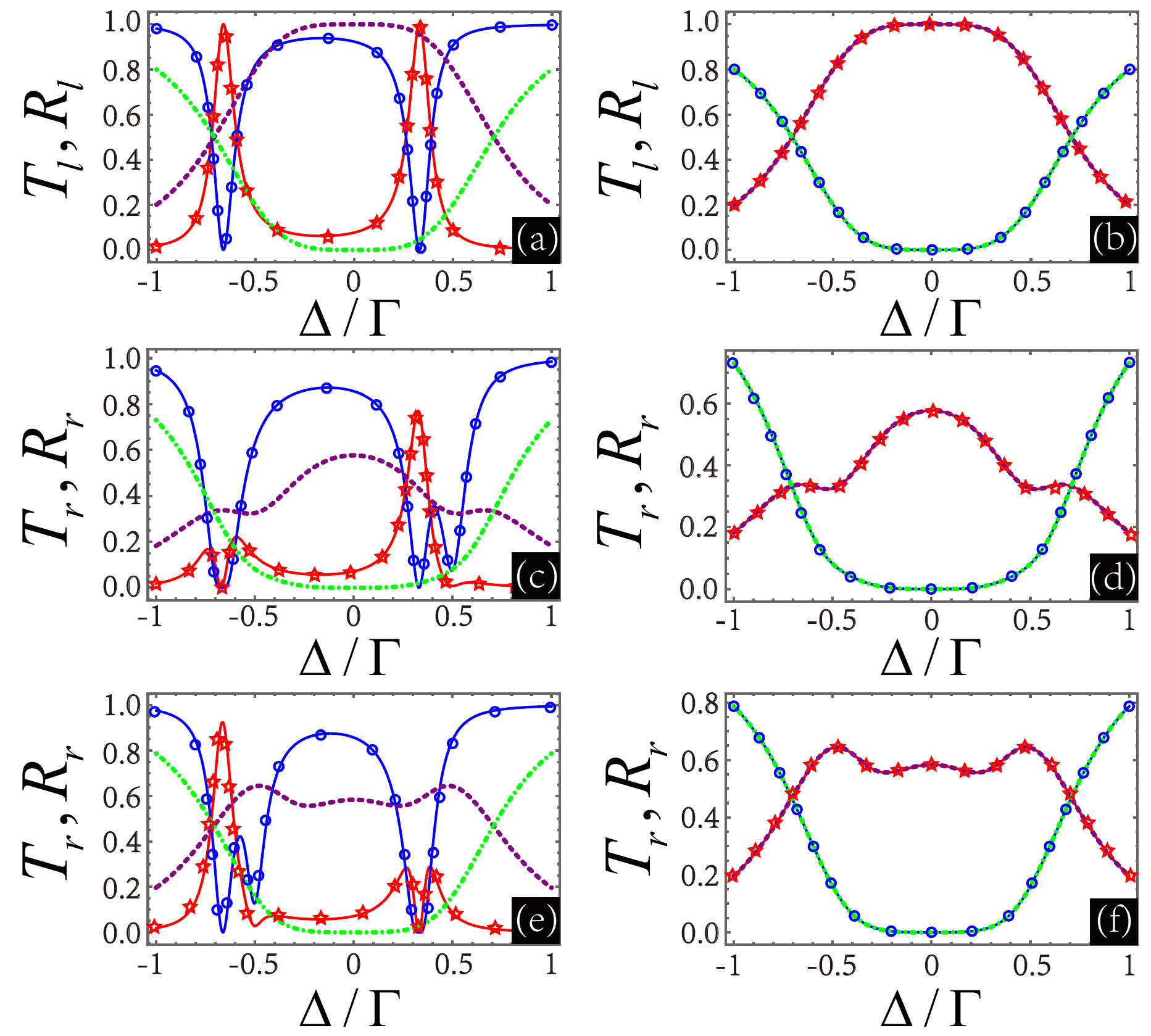}}
\caption{Comparisons of various scattering situations in quantum routers with and without Markovian approximations. (a)(b) show the transmission spectrum ${T_l}$ and reflection spectrum ${R_l}$ when the photon is input from the left side of the transmission line, while (c)-(f) describe the transmission and reflection ${T_r}$, ${R_r}$ of the right-incident photon. The purple dashed and green dotted-dashed lines describe the transmission spectrum ${T_r}$ and reflection spectrum ${R_r}$ under Markovian approximations, while the red star and blue circle lines represent the non-Markovian regimes. The parameters are chosen as ${\lambda _1} = {\lambda _2} = 0.95\Gamma$ and ${\lambda _3} = {\lambda _4} =\Gamma$ for (a)(c)(e), but ${\lambda _1} = {\lambda _2} = 999.5\Gamma$ and ${\lambda _3} = {\lambda _4} =1000 \Gamma$ for (b)(d)(f). The other parameters are the same as those in Fig.~\ref{fig5}.}\label{fig8}
\end{figure}

\begin{figure}[t]
\centerline{
\includegraphics[width=8cm, height=6cm, clip]{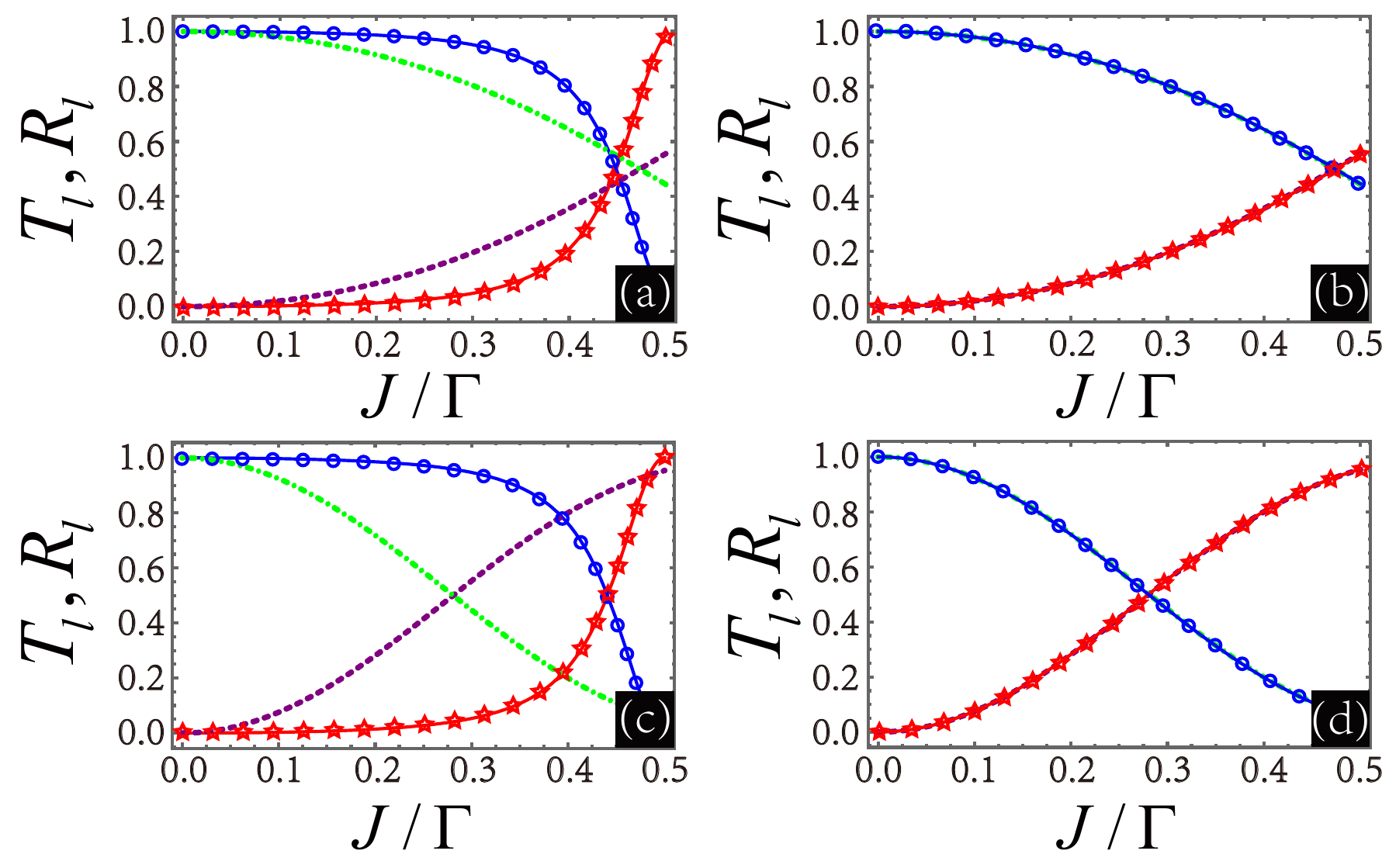}}
\caption{Comparisons of various scattering situations in quantum routers with and without Markovian approximations as a function of the hopping rate $J/\Gamma$ when the photon is input from the left side of the transmission line. The purple dashed and green dotted-dashed lines respectively represent the transmission and reflection spectra under Markovian approximations, while the red star and blue circle lines correspond to the cases in the non-Markovian regimes. To be specific, the parameters are chosen as ${\lambda _1} = {\lambda _2} = 0.95\Gamma$ and ${\lambda _3} = {\lambda _4} =\Gamma$ for (a)(c), but ${\lambda _1} = {\lambda _2} = 999.5\Gamma$ and ${\lambda _3} = {\lambda _4} =1000 \Gamma$ for (b)(d). The other parameters are the same as those in Fig. \ref{fig3}(c).}\label{fig9}
\end{figure}

In this section, we discuss the scattering behavior under Markovian approximations and compare it with that in the non-Markovian regimes. In the wide-band limit (i.e., ${\lambda _n} \to \infty $), the spectral density approximately takes ${J_n}(\omega ) \to {\Gamma _n}/2\pi $, equivalently, the spectral response functions $g(\omega ) \to \sqrt {{\Gamma _1}/2\pi } $, $G(\omega ) \to \sqrt {{\Gamma _2}/2\pi } $, $A(\omega ) \to \sqrt {{\Gamma _3}/2\pi } $, and $B(\omega ) \to \sqrt {{\Gamma _4}/2\pi } $. These describe the case in the Markovian limit. In many physical systems, the Markovian approximations are also valid when the coupling between the system and environments is weak and the characteristic time of the bath is sufficiently shorter than that of the system, i.e., ${\lambda _n} \gg {\Gamma _n}$ \cite{Breuer021002,Breuer210401}. Under the Markovian approximations, according to Eqs.~(\ref{k1234continuum}) and (\ref{correlation}), we obtain
\begin{equation}
\begin{aligned}
{h_1}(t) &=  - i\sqrt {{\Gamma _1}} \delta (t),\\
{h_2}(t) &= i\sqrt {{\Gamma _2}} \delta (t),\\
{h_3}(t) &=  - i\sqrt {{\Gamma _{3}}} \delta (t),\\
{h_4}(t) &= i\sqrt {{\Gamma _4}} \delta (t),\\
{f_n}(t) &= {\Gamma _n}\delta (t).
\label{kandf_Mar}
\end{aligned}
\end{equation}
Subsitituting Eq.~(\ref{kandf_Mar}) into Eq.~(\ref{Heisenberg_t>t0}), we can get
\begin{align}
\frac{d}{{dt}}{{p}_1}(t) &=  - (i{\omega _p} + \frac{{{\Gamma _1}}}{2} + {\gamma _p}){{p}_1}(t) - iJ{{p}_2}(t) - i\sqrt {{\Gamma _1}} p_{in}^{(1)}(t),\nonumber\\
\frac{d}{{dt}}{{p}_2}(t) &=  - (i{\omega _p} + \frac{{{\Gamma _2}}}{2} + {\gamma _p}){{p}_1}(t) - iJ{{p}_1}(t) + i\sqrt {{\Gamma _2}} p_{in}^{(2)}(t),\nonumber\\
\frac{d}{{dt}}{{q}_1}(t) &=  - (i{\omega _q} + \frac{{{\Gamma _3}}}{2} + {\gamma _q}){{q}_1}(t) - i\sqrt {{\Gamma _3}} q_{in}^{(1)}(t),\nonumber\\
\frac{d}{{dt}}{{q}_2}(t) &=  - (i{\omega _q} + \frac{{{\Gamma _4}}}{2} + {\gamma _q}){{q}_2}(t) + i\sqrt {{\Gamma _4}} q_{in}^{(2)}(t),\nonumber\\
\frac{d}{{dt}}m(t) &=  - (i{\omega _m} + {\gamma _m})m(t) - i\nu{{q}_2}(t).
\label{MarHeisenberg_t>t0}
\end{align}
Similarly, we can obtain the Markovian input-output relations by substituting Eq.~(\ref{kandf_Mar}) into Eq.~(\ref{input_output_relation}), which yield to
\begin{equation}
\begin{aligned}
p_{out}^{(1)}(t) - p_{in}^{(1)}(t) &=  - i\sqrt {{\Gamma _1}} {{p}_1}(t),\\
p_{out}^{(2)}(t) - p_{in}^{(2)}(t) &= i\sqrt {{\Gamma _2}} {{p}_2}(t),\\
q_{out}^{(1)}(t) - q_{in}^{(1)}(t) &=  - i\sqrt {{\Gamma _3}} {{q}_1}(t),\\
q_{out}^{(2)}(t) - q_{in}^{(2)}(t) &= i\sqrt {{\Gamma _4}} {{q}_2}(t).
\label{input_output_relation_Mar}
\end{aligned}
\end{equation}
It is worth mentioning that the Markovian input-output relations in Eq.~(\ref{input_output_relation_Mar}) are equivalent to those defined in Refs.\cite{Gardiner2000,Walls1994,Scully1997} and can return to the results of Refs.\cite{Gardiner2000,Walls1994,Scully1997} by the replacements ${g_k} \to i{g_k}$ $[g(\omega ) \to ig(\omega )]$, ${G_k} \to i{G_k}$ $[G(\omega ) \to iG(\omega )]$, ${A_k} \to i{A_k}$ $[A(\omega ) \to iA(\omega )]$, and ${B_k} \to i{B_k}$ $[B(\omega ) \to iB(\omega )]$ in Eqs.~(\ref{expand_solution}), (\ref{k1234continuum}), and (\ref{input_output_relation}).

In order to compare the results of the non-Markovian process with those of the Markovian process, we plot the evolution of the transmission and reflection spectra under this two situations in Fig.~\ref{fig8}. We find that when the environmental spectrum width $\lambda$ is small [such as Fig.~\ref{fig8}(a)], the so-called backflowing phenomenon occurs for the transmission and reflection spectra. From Fig.~\ref{fig8}(a), we can also see that in the Markovian cases, there is a small range of the detuning values which can result in transmission being $1$ and reflection being $0$ near $\Delta /\Gamma =0$, while in the non-Markovian cases, two peaks of the transmission (two valleys of the reflection) appear at $\Delta /\Gamma =-0.67$ and $\Delta /\Gamma =0.33$ respectively. For Fig.~\ref{fig8}(c) and (e), we have similar discussions. However, it is worth mentioning that for the case of right-input photon in the non-Markovian regimes (see Fig.~\ref{fig8}(c)(e)), we numerically solve the corresponding two detuning positions of total absorption of the scattering spectra based on the detuning of two peaks of transmission (i.e., $\Delta /\Gamma =-0.67$ and $\Delta /\Gamma =0.33$) in Fig.~\ref{fig8}(a). Therefore, given the same parameters, the scattering spectra at the corresponding detuning positions in Markovian approximations cannot achieve fully absorption. This indicates that the results in the non-Markovian regime are completely different from those under the Markovian approximation \cite{Ren013711}. As spectrum width $\lambda$ increases, the results given by the non-Markovian equation (\ref{Heisenberg_t>t0}) gradually tend to the Markovian ones \cite{Ren013711} given by Eq.~(\ref{MarHeisenberg_t>t0}), which result in the scattering spectrum in the non-Markovian regimes are in good agreement with those given in the non-Markovian limit, as shown in Fig.~\ref{fig8}(b)(d)(f).

Furthermore, we also compare the transmission and reflection spectra of the left-input photon as a function of the hopping rate $J /\Gamma$ in both Markovian and non-Markovian cases in Fig.~\ref{fig9}. Figure~\ref{fig9}(a)(b) describes the result of the first detuning position of transmission peak (reflection valley) in Fig.~\ref{fig8}(a), while Fig.~\ref{fig9}(c)(d) corresponds to the situation of the second detuning position of transmission peak (reflection valley) as shown in Fig.~\ref{fig8}(a). We find that the scattering behaviors in non-Markovian regimes are significantly different from the cases under Markovian approximations (see Fig.~\ref{fig9}(a)(c)). However, when environmental spectrum width $\lambda$ is large enough, i.e., in the non-Markovian limit, the scattering behaviors have a good agreement with the cases under Markovian approximations (see Fig.~\ref{fig9}(b)(d)), which is consistent with our expectations.

\section{Conclusion} \label{section V}
In summary, we have investigated a potentially practical scheme for the controllable single-photon transport via a cascaded quantum system, which is composed of two quantum nodes (two directly coupled CPW resonators and a superconducting ring resonator coupled to a single YIG disk) interconnected through a one-dimensional transmission line. The nonreciprocal single-photon routing properties in our model were studied, where the nonreciprocity is caused by the selective interaction between the magnon mode and the propagating microwave modes that share the identical chirality. We numerically and analytically studied the influences of the non-Markovian environments on the quantum router. The scattering behaviors of the system both under Markovian approximations and non-Markovian regimes were also described. The result showed that there is a small range of the detuning values which can cause transmission to be $1$ and reflection to be $0$ near $\Delta /\Gamma =0$ in the Markovian cases, while two peaks of the transmission (two valleys of the reflection) appear at $\Delta /\Gamma =-0.67$ and $\Delta /\Gamma =0.33$ respectively in the non-Markovian cases when the photon is input from the left side of the transmission. Corresponding to the positions of these two peaks, we numerically simulated the scattering behavior of the right-input photon under appropriate parameters and found that efficient routing can still be achieved in non-Markovian environments. Furthermore, we introduced the contrast ratio to quantitatively describe the nonreciprocity of the proposed nonreciprocal single-photon router and briefly discussd the possibility of experimental implementation. The corresponding experimental implementation of the model system is possible and feasible.

Quantum information processing \cite{Rohde052332,Li064083} and quantum network \cite{vanEnk4293,Clark177901,Yao318,Hong052302,Reiserer041003} are currently very popular fields of research. The results presented in this paper can protect the signal source from extraneous noise and has potential applications for the quantum information and quantum communication, which makes it possible to better understand the relations between the quantum routings and non-Markovianities. As a perspective, investigating the quantum routing beyond the rotating-wave approximation will be intriguing, such as the non-rotating-wave interactions between the resonators and environments $\sum\nolimits_k {{J_k}} (\hat c + {{\hat c}^\dag })({{\hat b}_k} + \hat b_k^\dag )$ \cite{Shen033805042121,Chen033603,Chen752024}, anisotropic non-rotating-wave interaction $\sum\nolimits_k {[{\xi _k}} (\hat b_k^\dag \hat c + {{\hat b}_k}{{\hat c}^\dag }) + {\zeta _k}({{\hat b}_k}\hat c + \hat b_k^\dag {{\hat c}^\dag })]$ \cite{Xie021046,Chen043708,Nakajima363,Ai042116,Zheng559}, and even the case which is not limited to the anisotropic non-rotating-wave approximation, i.e., all coupling including the resonators and environments which might be of the form $\sum\nolimits_{m,k} {({g_{mk}}{{\hat Y}_m}\hat Z_k^\dag  + g_{mk}^ * \hat Y_m^\dag {{\hat Z}_k} + {j_{mk}}{{\hat Y}_m}{{\hat Z}_k} + j_{mk}^ * \hat Y_m^\dag \hat Z_k^\dag )} $, where ${\hat Y_m^\dag }$ (${{{\hat Y}_m}}$) and ${\hat Z_k^\dag }$ (${{{\hat Z}_k}}$) represent the creation (annihilation) operators for the resonators, while ${{g_{mk}}}$ and ${{j_{mk}}}$ respectively denote the coupling strengths of the rotating-wave and non-rotating-wave interactions. The results might also be extended to possible applications including manipulating photon transport and designing reliable quantum networks and quantum communication protocols with non-Markovian effects.

\section*{ACKNOWLEDGMENTS}

H. Z. S. acknowledges National Natural Science Foundation of China under Grants No.~12274064 and Scientific Research Project for Department of Education of Jilin Province under Grant No.~JJKH20241410KJ. C. S. acknowledges financial support from the China Scholarship Council, the Japanese Government (Monbukagakusho-MEXT) Scholarship (Grant No.~211501), the RIKEN Junior Research Associate Program, and the Hakubi Projects of RIKEN. Y. H. Z. acknowledges the National Natural Science Foundation of China (NSFC)(Grants Nos.~12374333) and Jiangxi Provincial Natural Science Foundation (Grants Nos. 20212BAB201025).

\section*{}
\appendix*
\section{Discussions on the inhomogeneous term in Eq.~(\ref{elements})} \label{APPA}
\begin{figure*}[t]
\centerline{
\includegraphics[width=15.8cm, height=4.5cm, clip]{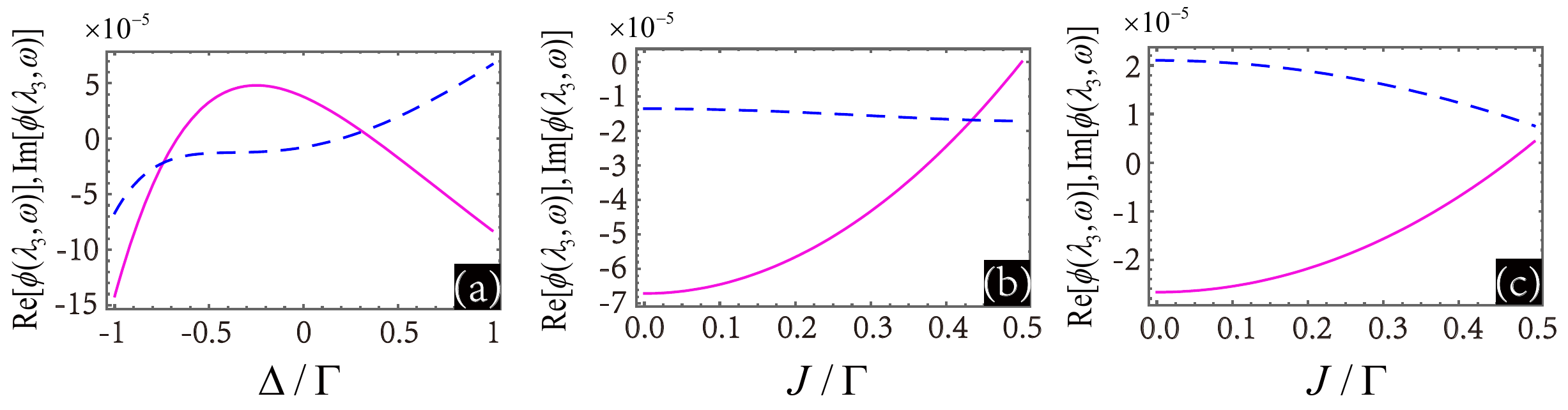}}
\caption{The real and imaginary parts of ${\phi _1}({\lambda _3},\omega )$ in the inhomogeneous term ${\varphi _1} = {t_1}{\phi _1}({\lambda _3},\omega )$ as a function of the detuning $\Delta /\Gamma $ for (a) and hopping rate $J/\Gamma $ for (b)(c) when the single photon is considered to be input from the left side of the transmission line. The magenta solid and blue dashed lines represent the real and imaginary parts of ${\phi _1}({\lambda _3},\omega )$, respectively. The parameters are chosen as $\omega=2.5 \Gamma$, $\theta=2 \pi$, ${\Gamma_1}={\Gamma_3}=\Gamma$, ${\gamma_p}=0$, ${\lambda_1}=0.95 \Gamma$, and ${\lambda_3}=\Gamma$, while $J=0.5 \Gamma$ for (a), ${\Delta}=-0.67 \Gamma$ for (b), and ${\Delta}=0.33 \Gamma$ for (c), where ${\Delta_p}=\Delta$.}\label{fig11}
\end{figure*}

The influences of inhomogeneous terms on the scattering behaviors should be discussed in two situations. 

Firstly, we consider the case where the single photon is input from the left side of the transmission line. By performing a simple substitution on the first equation of Eq.~(\ref{elements}), we can obtain ${t_{pq}} = {{\rm{exp}}({i\theta })}(1 - {t_1} + {\varphi _1}) ({t_2} - {\varphi _2})$ with ${t_1} = {{\tilde k}_3}(\omega ){h_3}( - \omega )/(i{\Delta _q} + {f_3}(\omega ) + {\gamma _q})$, ${t_2} = iJ{{\tilde k}_1}(\omega ){h_2}( - \omega )/[(i{\Delta _p} + {f_1}(\omega ) + {\gamma _p})(i{\Delta _p} + {f_2}(\omega ) + {\gamma _p}) + {J^2}]$, ${\varphi _1} = {t_1}{\phi _1}({\lambda _3},\omega )$, and ${\varphi _2} = {t_2}{\phi _2}({\lambda _1},\omega )$, where ${\phi _1({\lambda _3},\omega )} = q_{in}^{(1)}(i{\lambda _3})/q_{in}^{(1)}(\omega )$ and ${\phi _2({\lambda _1},\omega )} = p_{in}^{(1)}(i{\lambda _1})/p_{in}^{(1)}(\omega )$. Similarly, for the second equation of Eq.~(\ref{elements}), we have ${r_{pp}} = 1 - {r_1} + {\varphi _3}$, where ${r_1} = (i{\Delta _p} + {f_2}(\omega ) + {\gamma _p}){{\tilde k}_1}(\omega ){h_1}( - \omega )/[(i{\Delta _p} + {f_1}(\omega ) + {\gamma _p})(i{\Delta _p} + {f_2}(\omega ) + {\gamma _p}) + {J^2}]$ and ${\varphi _3} = {r_1}{\phi _2}({\lambda _1},\omega )$. 

In Markovian approximations, ${\phi _1}({\lambda _3},\omega )$ and ${\phi _2}({\lambda _1},\omega )$ tend to zero for ${\lambda _1},{\lambda _3} \to \infty $. We find that ${\varphi _1} = {t_1}{\phi _1}({\lambda _3},\omega )$, ${\varphi _2} = {t_2}{\phi _2}({\lambda _1},\omega )$, and ${\varphi _3} = {r_1}{\phi _2}({\lambda _1},\omega )$ are induced by non-Markovian effects and have no Markovian counterparts, which are inhomogeneous terms depending on the specific forms of the input field $q_{in}^{(1)}(t)$ and $p_{in}^{(1)}(t)$, respectively. In order to investigate the effect of the inhomogeneous terms, we now assume that the input field $p_{in}^{(1)}(t)$ has the form of damped oscillation $p_{in}^{(1)}(t) = {a_1}{e^{ - \gamma t}}\sin ({b_1}{t^2})$ for $\gamma > 0$ and ${b_1}>0$, which yields 
\begin{widetext}
\begin{align}
&{\phi _2}\left( {{\lambda _1},\omega } \right) = \frac{{\cos [\frac{{{{(\gamma  + {\lambda _1})}^2}}}{{4{b_1}}}][1 - 2fc(\frac{{\gamma  + {\lambda _1}}}{{\sqrt {2\pi {b_1}} }})] + [1 - 2fs(\frac{{\gamma  + {\lambda _1}}}{{\sqrt {2\pi {b_1}} }})]\sin [\frac{{{{(\gamma  + {\lambda _1})}^2}}}{{4{b_1}}}]}}{{\cos [\frac{{{{(\gamma  - i\omega )}^2}}}{{4{b_1}}}][1 - 2fc(\frac{{\gamma  - i\omega }}{{\sqrt {2\pi {b_1}} }})] + [1 - 2fs(\frac{{\gamma  - i\omega }}{{\sqrt {2\pi {b_1}} }})]\sin [\frac{{{{(\gamma  - i\omega )}^2}}}{{4{b_1}}}]}},\\ 
&{\rm{where}} \ fs(z) = \int_0^z {\sin } \left( {\pi {t^2}/2} \right)dt \quad {\rm{and}} \quad  fc(z) = \int_0^z {\cos } \left( {\pi {t^2}/2} \right)dt. \nonumber
\end{align}
\end{widetext}
According to the non-Markovian input-output relations in the frequency domain given by Eq.~(\ref{input_output_relation_fre}), we can easily derive $q_{in}^{(1)}(\omega ) = {p_2}(\omega ){h_2}( - \omega ){e^{i\theta }}$, where ${p_2}(\omega )$ can be obtained by solving the quantum Langevin equation in the frequency domain given by Eq.~(\ref{Heisenberg_fre}). After a cumbersome but direct calculation, we find ${\phi _1}({\lambda _3},\omega )$ is a function related to the hopping rate $J$ and detuning ${\Delta}$, where the real (magenta solid lines) and imaginary (purple dashed lines) parts of ${\phi _1}$ versus ${\Delta/\Gamma}$ and $J/\Gamma$ are plotted in Fig.~\ref{fig11}. From Fig.~\ref{fig11}, we can see that both the real and imaginary parts of ${\phi _1}({\lambda _3},\omega )$ reach the order of ${10^{ - 5}}$ within the parameter range we selected, that is to say, $| {{\phi _1}({\lambda _3},\omega )} |$ is small enough, indicating that the inhomogeneous terms $|{\varphi _1}|$ and $|{\varphi _2}|$ are much smaller than $|{t_1}|$ and $|{t_2}|$. For ${\phi _2}({\lambda _1},\omega )$, we can directly derive $| {{\phi _2}({\lambda _2},\omega )} | \sim  {10^{ - 5}}$. Therefore, the influences of the inhomogeneous terms $|{\varphi _1}|$ and $|{\varphi _2}|$ on the cavity transmission and reflection can be ignored.

Similarly, for the case of the right-inputting photon, we can also have the discussions above. For convenience, we take $q_{in}^{(2)}(t)$ with the form of damped oscillation above, which yields
\begin{widetext}
\begin{align}
   &{\phi _3}\left( {{\lambda _4},\omega } \right) = \frac{{\cos [\frac{{{{(\gamma  + {\lambda _4})}^2}}}{{4{b_1}}}][1 - 2fc(\frac{{\gamma  + {\lambda _4}}}{{\sqrt {2\pi {b_1}} }})] + [1 - 2fs(\frac{{\gamma  + {\lambda _4}}}{{\sqrt {2\pi {b_1}} }})]\sin [\frac{{{{(\gamma  + {\lambda _4})}^2}}}{{4{b_1}}}]}}{{\cos [\frac{{{{(\gamma  - i\omega )}^2}}}{{4{b_1}}}][1 - 2fc(\frac{{\gamma  - i\omega }}{{\sqrt {2\pi {b_1}} }})] + [1 - 2fs(\frac{{\gamma  - i\omega }}{{\sqrt {2\pi {b_1}} }})]\sin [\frac{{{{(\gamma  - i\omega )}^2}}}{{4{b_1}}}]}},\\
   &\text{where} \ fc(z) = \int_0^z {\cos } \left( {\pi {t^2}/2} \right)dt \quad \text{and} \quad fs(z) = \int_0^z {\sin } \left( {\pi {t^2}/2} \right)dt. \nonumber
\end{align}
\end{widetext}
Based on the non-Markovian input-output relations in Eq.~(\ref{input_output_relation_fre}), we can obtain $p_{in}^{(2)}(\omega ) = [q_{in}^{(2)}(\omega ) + {q_2}(\omega ){h_4}( - \omega )]{e^{i\theta }}$. The third and fourth equations of Eq.~(\ref{elements}) can also be respectively rewritten as 
\begin{widetext}
\begin{align}
{t_{qp}} &= {{\rm{exp}}({i\theta })}[{t_3} - {t_3}{\phi _3}({\lambda _2},\omega )] [1 + {t_4}{\phi _4}({\lambda _4},\omega ) - {t_4}],\nonumber\\
{r_{qq}} &= {{{\rm{exp}}({2i\theta })}}[1 - {r_2} + {r_2}{\phi _1}({\lambda _3},\omega )] [1 + {r_3} - {r_3}{\phi _3}({\lambda _2},\omega )][1 + {r_4}{\phi _4}({\lambda _4},\omega ) - {r_4}],\nonumber\\
{t_3} &= iJ{{\tilde k}_2}(\omega ){h_1}( - \omega )/[(i{\Delta _p} + {f_1}(\omega ) + {\gamma _p})(i{\Delta _p} + {f_2}(\omega ) + {\gamma _p}) + {J^2}],\nonumber\\
{t_4} &= (i{\Delta _m} + {\gamma _m}){{\tilde k}_4}(\omega ){h_4}( - \omega )/[(i{\Delta _q} + {f_4}(\omega ) + {\gamma _q})(i{\Delta _m} + {\gamma _m}) + {\nu ^2}],\nonumber\\
{r_2} &= {{\tilde k}_3}(\omega ){h_3}( - \omega )/[i{\Delta _q} + {f_3}(\omega ) + {\gamma _q}],\nonumber\\
{r_3} &= {{\tilde k}_2}(\omega )(i{\Delta _p} + {f_2}(\omega ) + {\gamma _p}){h_2}( - \omega )/[(i{\Delta _p} + {f_1}(\omega ) + {\gamma _p})(i{\Delta _p} + {f_2}(\omega ) + {\gamma _p}) + {J^2}],\nonumber\\
{r_4} &= (i{\Delta _m} + {\gamma _m}){{\tilde k}_4}(\omega ){h_4}( - \omega )/[(i{\Delta _q} + {f_4}(\omega ) + {\gamma _q})(i{\Delta _m} + {\gamma _m}) + {\nu ^2}],
\end{align}
\end{widetext}
with ${\phi _3}({\lambda _2},\omega ) = p_{in}^{(2)}(i{\lambda _2})/p_{in}^{(2)}(\omega )$ and ${\phi _4}({\lambda _4},\omega ) = q_{in}^{(2)}(i{\lambda _4})/q_{in}^{(2)}(\omega )$. After calculations, we prove $|{\phi _3}({\lambda _2},\omega )|$ and $|{\phi _4}({\lambda _4},\omega )|$ are small enough, which also indicates that the influences of the inhomogeneous terms on the cavity transmission and reflection can be ignored for plotting.

\end{document}